# The Premartensite and Martensite in The Fe$_{50}$Rh$_{50}$ System

Esmaeil Adabifiroozjaei[1*], Fernando Maccari[2], Lukas Schäfer[2], Tianshu Jiang[1], Oscar Recalde-Benitez[1], Alisa Chirkova[2,3], Navid Shayanfar[2], Imants Dirba[2], Nagaarjhuna A Kani[1,4], Olga Shuleshova[5], Robert Winkler[1], Alexander Zintler[1,6], Ziyuan Rao[7], Lukas Pfeuffer[2], András Kovács[8], Rafal E. Dunin-Borkowski[8], Michael Farle[9], Konstantin Skokov[2], Baptiste Gault[7,10], Markus Gruner[9], Oliver Gutfleisch[2], and Leopoldo Molina Luna[1]

[1] Advanced Electron Microscopy, Institute of Material Science, Technical University of Darmstadt, 64287 Darmstadt, Germany
[2] Functional Materials, Institute of Materials Science, Technical University Darmstadt, 64287 Darmstadt, Germany
[3] Bielefeld Institute for Applied Materials Research, Bielefeld University of Applied Sciences, D-33619 Bielefeld, Germany
[4] Division of Research with Neutrons and Muons, Paul Scherrer Institute, Switzerland
[5] The Leibniz Institute for Solid State and Materials Research, 01069, Dresden, Germany
[6] Karlsruhe Institute of Technology, Laboratory for Electron Microscopy (LEM), Engesserstr. 7, 76131 Karlsruhe, Germany
[7] Max-Planck-Institut für Eisenforschung, Düsseldorf 40237, Germany
[8] Ernst Ruska-Centre for Microscopy and Spectroscopy with Electrons and Peter Grünberg Institute, Forschungszentrum Jülich, Jülich 52425, Germany
[9] Faculty of Physics and Center for Nanointegration Duisburg–Essen (CENIDE), University of Duisburg–Essen, Duisburg 47057, Germany
[10] Department of Materials, Royal School of Mines, Imperial College London, London, SW7 2AZ, UK

* Corresponding author: e.adabifiroozjaei@aem.tu-darmstadt.de

## *Abstract*

Metallic/intermetalic materials with BCC structures hold an intrinsic instability due to phonon softening along [110] dirrection, causing BCC to lower-symmetry phases transformation when the BCC structures are thermally or mechanically stressed. Fe$_{50}$Rh$_{50}$ binary system is one of the exceptional BCC structures (ordered-B2) that has not been yet showing such transformation upon application of thermal stress, although mechanical deformation results in B2 to disordered FCC (γ) and L10 phases transformation. Here, a comprehensive transmission electron microscopy (TEM) study is conducted on thermally-stressed samples of Fe$_{50}$Rh$_{50}$ aged at water and liquid nitrogen from 1150 °C and 1250 °C. The results show that, samples quenched from 1150 °C into water and liquid nitrogen show presence of 1/4{110} and 1/2{110} satellite reflections, the latter of which is expected from phonon dispersion curves obtained by density functional theory calculation. Therefore, it is believed that Fe$_{50}$Rh$_{50}$ maintains the B2 structure that is in premartensite state. Once Fe$_{50}$Rh$_{50}$ is quenched from 1250 °C into liquid nitrogen, formation of two short-range ordered tetragonal phases with various *c/a* ratios (~1.15 and 1.4) is observed in line with phases formed from mechanically deformed (30%) sample. According to our observations, an accurate atomistic shear model ({110}<1-10>) is presented that describes the martensitic transformation of B2 to these tetragonal phases. These findings offer implications useful for understanding of magnetic and physical characteristics of metallic/intermetallic materials.

**Introduction**

FeRh alloys (Fe$_{48}$Rh$_{52}$ to Fe$_{56}$Rh$_{44}$) have ordered-B2 structure (cubic CsCl-prototype) with an antiferromagnetic (AFM) to ferromagnetic (FM) transition near room temperature [1]. The transition is isostructural with about 1% change in volume and is accompanied by giant magnetoresistive and magnetocaloric effects [2]. Therefore, there has been a lot of interest on these alloys (particularly as thin film) for technological applications like heat assisted magnetic recording (HAMR) [3], antiferromagnetic spintronics [4], and magnetic refrigeration [5,6]. Additionally, there has been particular interest on Fe$_{50}$Rh$_{50}$ as a model system to investigate the effect of external fields on the first order transitions [7].

Since the transition is isostructural, the majority of previous studies did not focus on the local atomic structure characterization of the alloys in either AFM or FM states. The studies that have been conducted so far usually provide an average picture of the structure determined applying high-resolution X-ray diffraction [8], X-ray magnetic circular dichroism [9], X-ray photoelectron spectroscopy, and nuclear resonant inelastic X-ray scattering techniques [2]. Although these techniques benefit from high accuracy, they cannot visualize the local atomic structure and correlated defect structures of FeRh alloy. To this end, scanning transmission electron microscopy (S/TEM) with ability to resolve localized atomic structure could be utilized, as used [10,11].

Recent detailed TEM studies of FeRh alloys conducted by McLaren [1] evidenced of ordered modulations, when the alloys are observed along [100], [110], and [111] directions. Independent studies also reported similar and unique superlattice reflections in the [110] zone axis [12,13] in thin films and bulk alloys.

On the other hand, extensive structural characterization of similar systems (Zr [14], Ti [15], Ni-Al [16], CdAu [17], TiN [18], CuZn [19], CuALN2i [19], Ni$_2$FeGa [20] *etc…*, all having BCC structure) showed presence of premartensite structure before martensite transformation. The premartensite structure manifests itself with usually non-commensurate superlattice reflections in the reciprocal space, while in real space it appears as ordered modulations. According to Zener's theory [21], the ordered-B2 structures (called also β-series of alloys) show softening in elastic constant C' (corresponding to {110}⟨1-10⟩ shear modulus) when the temperature is lowered. Such softening in C' yields an anisotropic factor (A=C'/C$_{44}$) leading to an ordered displacement of atoms along [110] and consequent appearance of ordered modulations [22]. In order to see if such softening occurs in FeRh alloy or not, comparison of elastic constants is necessary. However, experimental data is not available in the literature for FeRh alloys since single crystal of FeRh are extremely difficult to grow. Yet, there has been a recent density functional theory (DFT) calculation on the elastic constants of FeRh alloy, from which the A is 5.4 for AFM state [23]. This can be compared with similar alloys (NiTi (2), NiAl (9), AuCd (14-11)) [22], implying that from the anisotropy in elastic constant point of view a premartensite structure is expected in FeRh system.

However, to the authors' knowledge, there have been no reports of premartensite structure in FeRh system. Nonetheless, when FeRh is alloyed with Pd [24] or Pt [25], ordered tetragonal structures (L10) appear. Furthermore, recent detailed DFT calculations made it clear that B2 structure in AFM state is not the ground state in the FeRh system [2,8,23,26–28]. In fact, it was shown that there are imaginary soft modes in the phonon dispersion band of ordered-B2 and when they are fully relaxed, lower symmetry structures such as orthorhombic or monoclinic are obtained. Although monoclinic structure was never observed, but the orthorhombic one was stabilized in epitaxially strained film of FeRh on tungsten substrate [8]. As the authors mentioned, the magnetic properties of such disordered orthorhombic structure are deteriorated relative to B2. Additionally, it is also known that electrically-induced strain can control the AFM to FM transition in a reversible manner [9].

Apart from the effect of epitaxial strain on the structure and property of FeRh system, it is also known that plastic deformation of FeRh alloy can induce B2 to L10 (c/a~1.15) and γ transformation, the latter of which is known to be disordered and displaying spin-glass-like behaviour [12,29–31]. In contrast, γ obtained from dealloying process is in ferromagnetic state at room temperature [32]. The γ is known to transform back to B2 upon heating at ~150 °C or higher [30,33]. Interestingly the properties of B2 structure (AFM to FM transformation) can also be recovered upon the heating. Since diffusion is limited at such low temperature, it is not really known how a disordered phase can be transformed to an ordered phase (B2). Yavari suggested that L10 and γ have probably short-range order (SRO) and that is why they show spin-glass-like behaviour [30]. However, there has not been still a concrete description of transformation mechanism for mechanically stressed FeRh alloy. Additionally, yet it is not known how FeRh alloy respond to thermal stress, as it is known to also causes displacive transformation [34].

In order to shed light on the raised matters, the present study used S/TEM techniques to comprehensively study the local atomic structure of $Fe_{50}Rh_{50}$ alloys that were thermally stressed. Three types of thermally-stressed samples were examined: *(1)* quenched from 1150 °C into water, *(2)* quenched from 1150 °C into liquid nitrogen (LN2), *(3)* and quenched from 1250 °C into LN2. The latter is compared with a mechanically deformed (30% plastic deformation) sample. From the data acquired in samples 1 and 2, it was observed that there are satellite reflections at 1/4 and 1/2 of {110} plane reflections related to slight atomic displacement of both Fe and Rh along {110} planes. Additionally, our DFT calculation predicted phonon softening at reciprocal points >1/2[ζζ0]. Accordingly, it was concluded that these samples are in a premartensite state. In contrast, sample 3 showed two L10 phases with different *c/a* ratio (~1.15 and 1.4). These same phases formed in mechanically stressed sample. A shuffling mechanism was proposed that describes perfectly the B2 to these L10 structures transformation. Overall, our results indicate that FeRh B2 structure has an intrinsic instability which causes its transformation to lower symmetry structures induced by thermal or mechanical stress.

**Experimental Procedure**

The ingot with the composition $Fe_{50}Rh_{50}$ was prepared by arc melting of the appropriate amounts of pure Fe (99.995%, Alfa Aesar) and Rh (99.9%, Alfa Aesar) in a water-cooled Cu-crucible. The sample was remelted 3 times to ensure homogeneity, after which it was suction-cast in a 2-mm rod. The rod was cut to thin plates of the approximate thickness of 0.5 mm for subsequent heat treatment. The chemical composition of alloy was determined (Fe: 50.11, Rh: 49.89 at%) using inductively coupled plasma (ICP) spectroscopy. The samples were sealed in quartz tubes under pure Ar atmosphere (0.2 bar) and were subjected to annealing to obtain the ordered-B2 phase. The annealing was done at 1150°C for 72 hours followed by water quenching. The tube was not broken in order to prevent oxidation. Slow cooling was also tested from 1150°C with cooling rate of 0.1 degree/min. For thermal quenching in LN2, the samples were encapsulated in quartz tubes backfilled with Ar and quenched from 1150°C and 1250°C (with dwelling time of 3 hours) into LN2. In this process the tube was broken in order to let the alloy quickly reach LN2 temperature. Afterwards, the surface of the samples was finely polished and room temperature X-ray diffraction was recorded using a Bruker D8 advance diffractometer in Bragg-Brentano geometry with Cu K_alpha radiation and energy dispersive detector. A step size of 0.02° was used and the sample was continuously rotated during the measurement. Thermomagnetic measurements were carried out using a vibrating sample magnetometer (VSM) in a magnetic property measurement system (MPMS3, Quantum Design USA) under applied field ($\mu_0H_{app}$) of 1 T. Heating and cooling curves were recorded within a temperature range between 10 K and 450 K with temperature rate of 2 K/min. Mechanical deformation was done using Instron Universal Test Machine.

For TEM investigations, electron transparent thin sections were prepared using various methods. Most of the lamella were prepared conventionally using mechanical polishing for initial thinning and subsequent Argon milling by precision ion polishing system (PIPS: Gatan Dual Ion Mill Model 600) to create small hole in the sample. For initial PIPS process, an angle of 4° and energy of 4.5 KeV were used. Final milling was done using angle of 1.5° and energy of 2 keV. Milling was done at LN2 temperature. Focused ion beam (FIB: FIB: JEOL JIB-4600F) milling using Ga ion also was used. Initial and final milling energies were set to 30 kV and 2 kV, respectively. Regardless of the method to prepare lamella, similar nanostructural features were observed in FeRh alloy.

The TEM investigation was done using JEM-2100, JEOL machine while STEM observations were made by aberration corrected JEM-ARM200, JEOL, FEI Titan 80-200 (ChemiSTEM) and FEI Titan Tecnai G2 F20 machines. In TEM mode, bright-field, dark-field, and high resolution TEM images were aqcuired, while selected area diffraction pattern was used to characterize the structure. In STEM mode, high-angle annular dark-field (HAADF) images were recorded using a detector semi-angle of 90-370 mrad. STEM image simulations in this study were performed using multislice algorithm via prismatic [35,36] software. The probe size used for the simulations was 0.5 Å × 0.5 Å, and the semi-angle was set to 30 mrad. To obtain the simulated HAADF-STEM images, we selected the inner and outer virtual detectors as 90 mrad and 125 mrad, respectively. The diffraction pattern simulation was done using CrysTBox (Crystallographic Toolbox). The visualization of crystal structure was performed using VESTA (Visualization for Electronic and Structural Analysis) [37]. The nano-scale elemental distributions in the alloys were investigated by atom probe tomography (APT) (LEAP 5108XR, Cameca Inc.) at a pulse repetition rate of 200 kHz and a pulse energy of 80 pJ. The operating temperature was 60 K and the target detection rate was set to five ions detected every 1000 pulses. Site-specific lift-out of APT tips was performed from the homogenized and annealed alloys with a FIB instrument (an FEI Helios Plasma focused ion beam (PFIB)) [38–40].

The calculation of the phonon dispersion relations were carried out as described in [2] within the so-called direct or approach using the PHON code [41] to diagonalize the dynamical matrix. Force constants were obtained from density functional theory (DFT) calculations involving ionic displacements of 0.02 Å in a 3×3×3 supercell constructed from the 4-atom primitive cell with fcc basis describing the antiferromagnetic ground state structure. For this step, we employed the plane wave code VASP [42,43] in combination with the generalized gradient approximation for exchange and correlation in the formulation of Perdew, Burke and Ernzerhof [44]. We use projector augmented wave [43] potentials with 3p, 3d, and 4s valence states for Fe and 4p, 4d, and 5s for Rh together with a plane wave cutoff $E_{cut}$=450 eV. Brillouin zone integration was carried out on a 4×4×4 Monkhorst-Pack k-grid in combination with a finite temperature smearing according to Methfessel and Paxton [45] with a broadening of 0.1 eV.

**Results and Discussion**

**The premartensite ptructure**

*(1) Quenching from 1150 °C into water*

Figures 1a-c show the electron diffraction patterns of sample 1 along three principal zone axes ([001], [1-10], [-111]), while Figures 1d-f show the HRTEM images of the sample along the corresponding directions. The main reflections in all three zone axes match the B2 structure. However, there is a systematic presence of extra reflections (marked with dashed yellow circles) and in HRTEM corresponding modulations (marked with dashed yellow lines) along {100} and {110} planes reflections. The extra reflections can be inclined from the main reflections with angles smaller than 15° (Figures 1b and c) or sometimes aligned closely with the main reflections (Figure 1a and Figure S1). The spacing of the modulations, as shown in inset of Figures 1a and b are ~0.8 for (100) plane and 1.2 1/nm for (110) plane. These are located at ~1/4 of {100} and {110} planes reflections. The spacing of these modulations varies between 0.6-0.8 1/nm for modulation along {100}, while that for modulation along {110} varies in the range of 1.0-1.2 1/nm, meaning that they are incommensurate. These variations were consistently observed on different domains within the same sample and in different samples. Interestingly, each domain has different modulations as shown in Figure S2. The size of each domain can be as large as few hundred nanometres, while the grain size in the sample ranges between a few to tens of micrometre. Two previous studies reported extra reflections at [1-10] zone axes [12,13], as shown in Figure S3. Using dark-field imaging, Takahashi *et al.* [12] stated that the extra reflections are caused by fine nanoprecipitates of the γ phase formed during cold rolling, and Castiella *et al.* [13] supported this hypothesis. However, Castiella's thin films were annealed at 700 °C for 6 hours, sufficient for transforming γ to B2 [33]. Additionally, the spacings of these extra reflections do not match with those expected for the γ phase. Additionally, the X-ray pattern (Figure S4) of the present sample showed no trace of γ phase after quenching from 1150 °C into water, while in the as-cast state the presence of γ reflections is clear.

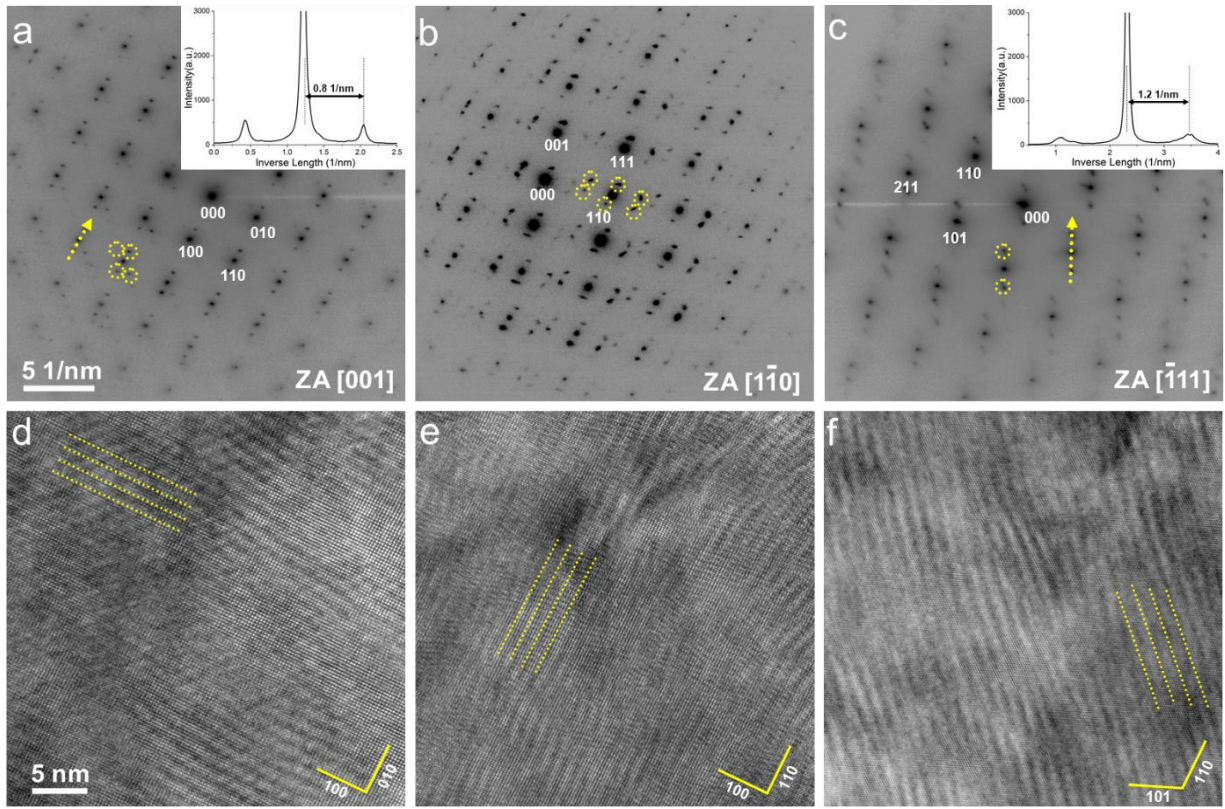

**Figure 1.** TEM electron diffraction patterns (**a**, **b**, **c**) and corresponding HRTEM images (**d**, **e**, **f**) of FeRh alloy in three principal zone axes. The dashed circles in (**a**,**b**,**c**) show the satellite reflections. The insets in **a** and **c** are intensity scan recorded from the dashed arrows showing the distance of modulations from main reflections. Dashed lines in **d**, **e**, and **f** highlight the modulations.

In order to examine the variation in chemical composition, atom probe tomography measurements were carried and the results are presented in Figure S5. Although there is slight variation in chemical composition as the 1D chemical composition shows (Figure S5d), but this cannot be related to the modulation observed in the sample as the frequency of the chemical composition change (~6 nm) is much longer than that observed in the sample (~ 1 nm). To further study the origin of the extra reflections dark-field (DF) TEM imaging was applied, revealing a modulated nanostructure as shown in Figures 2a and b. It worth noting that no trace of γ phase in the matrix of B2 ordered phase was found. In order to clarify if the quenching in water resulted in the modulated nanostructure, the same alloy was slowly cooled (0.1 degree/minute from 1150°C). The diffraction diffraction pattern in Figure S6. reveals modulations in the nanostructure. A comparison of the magnetic properties of both alloys, quenched in water and slow-cooled, was performed and the results are presented in Figure 2c. The transition temperature is slightly higher for the slow-cooled alloy than that quenched in water, but both samples show very sharp transition. These observations evidence that the extra reflections are finger print of FeRh alloy (at least for ~50/50 alloys) with a B2-ordered structure.

As mentioned before, such modulated structures were observed in similar alloys and they were ascribed to phonon softening in [110] directions. For example, in NiTi alloys [22] or stoichiometric $Ni_2MnGa$ [46], the phonon softening is manifested at ~1/3TA2[$\zeta\zeta$0] (transverse acoustical) and hence extra reflections appear at ~1/3 of {110} planes reflections in reciprocal spaces. Therefore, in $Fe_{50}Rh_{50}$ system, if the observed modulations are related to phonon softening then very low phonon frequencies or even minimums should be present at ~1/4TA2[$\zeta\zeta$0] and ~1/4TA1[$\zeta$00]. Unfortunately, there is only one experimental report of inelastic neutron spectroscopy measurement of FeRh, which is not very conclusive [47]. Hence, density functional theory (DFT) calculation was carried out to produce phonon dispersion curves for AFM state of FeRh. The phonon dispersion curves along [$\zeta\zeta$0] and [$\zeta$00] were obtained and given in Figure 2d. As seen, all branches start to show low frequencies or even minimums when $\zeta$ is larger than 1/2[$\zeta$00] or 1/2[$\zeta\zeta$0]. Additionally, the frequency for boundary of TA1[$\zeta$00] and TA2[$\zeta\zeta$0] zones approach negative values, implying that the B2 structure in AFM state is not stable. Another interesting fact is the very low frequency observed for the whole TA2[$\zeta\zeta$0] branch, while for $\zeta$ larger than 1/2 the value starts to become close to zero. From the current DFT data, one can infer that the B2 structure should show extra reflections at points larger than 1/2TA2[$\zeta\zeta$0] and ~1/2TA2[$\zeta$00], while the current TEM data present extra reflections at ~1/4TA2[$\zeta\zeta$0] and ~1/4TA1[$\zeta$00]. However, it has to be considered that DFT calculation is done at zero Kelvin, whereas the TEM data is collected at room temperature. Furthermore, in similar studies on Zr and Ti metals [14,48], very low frequencies (not minimum) were observed at ~1/2TA2[$\zeta\zeta$0]. This is known to be responsible for instability of BCC structure which finally causes displacement of atoms along [110] and turning BCC to hexagonal (HCP).

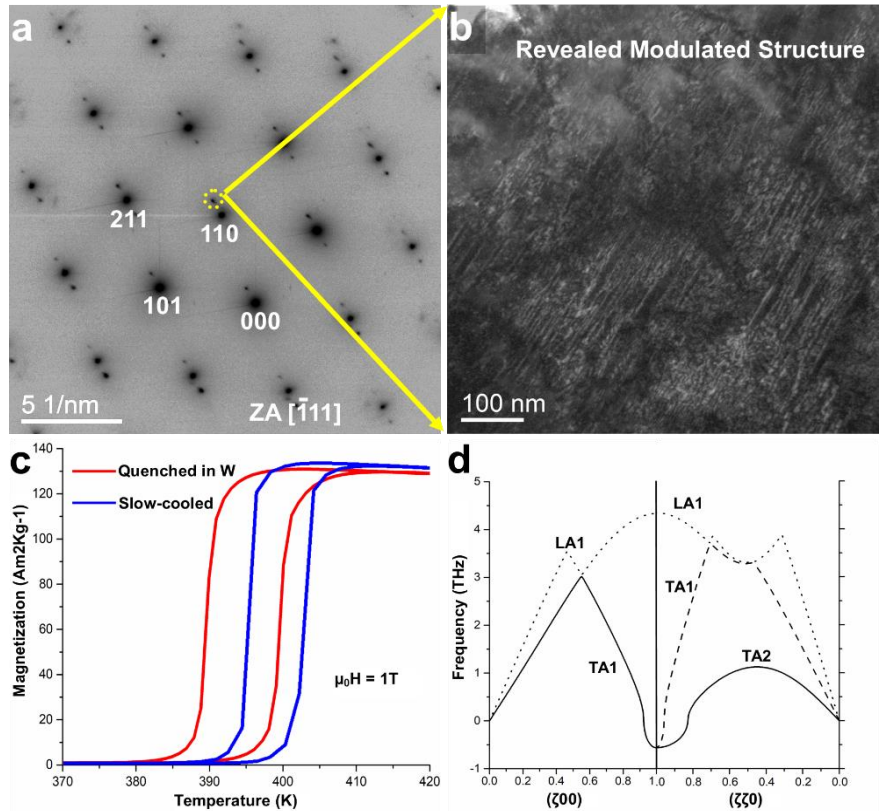

**Figure 2.** **a** Electron diffraction pattern taken in [-111] zone axis from sample 1 and **b** is a dark-filed image recorded using the extra satellite reflection circled by dashed yellow line in **a**. **c** Magnetization *vs* temperature measurement for alloys cooled through quenching in water (sample 1: red) and with 0.1 degree/min cooling rate (blue) from 1150 °C. **d** Phonon dispersion curves of $Fe_{50}Rh_{50}$ alloy (for AFM state) recorded along [ζζ0] and [ζ00] done at zero Kelvin.

To gain insight about the cause of the ~1/4{110} reflection, high resolution HAADF-STEM imaging along [001] was carried out. Figures 3a, b, and c are HAADF images of same area at different magnifications. Figures 3a (also the inset) and b show that the modulation is uniformly present in the nanostructure while it has slight deviation from (110) plane reflection. Figure 3c provides a closer look at the nanostructure, in which now it is possible to observe that in some portions of the nanostructure Fe and Rh atoms are smeared out relative to the other atoms and these are in fact the areas that form the modulated structure. To clarify this further, line profile along 5 lines in the Figure 3c were analysed. According to the B2-ordered structure, the atomic profile for line 1, 3, and 5 should be same, while same is true for lines 2 and 4. These line profiles are plotted in Figure 3e. While line profiles of 1 and 5 are both sharp and very alike, line 3 shows a smeared-out profile. On the other hand, line profiles of 2 and 4 reveal another interesting feature, which is a relative shift (0.3 Å) in Fe and Rh peaks. These results indicate that the modulated structure in the B2-ordered structure is due to displacement of atoms in (110) along [1-10], while the shift directions are different for different atomic rows. Based on the acquired information, a supercell with some atoms displaced about 0.3 Å is prepared (schematic of supercell is shown as Figure S7 and HAADF-STEM image simulation was conducted as shown in Figure 3d. In order to accurately evaluate the structure, 5 line-profiles were made like in Figure 3c and the results are shown as Figure 3f. A good agreement of line profiles in Figures 3e and f were obtained. Furthermore, theoretical diffraction pattern of the prepared supercell along [001] was simulated (see Figure S8). The resemblance of the simulated diffraction pattern and the FFT image (inset in Figure 3a) is very high. It should be also noted that the length of displacement, which determines the intensity of the extra reflection, is not uniform all over the sample. As an example, Figure S9 shows a HAADF-STEM image taken from another sample prepared along the [001] crystallographic direction. As the HAADF image and corresponding FFT image show, the modulation is not easily visible.

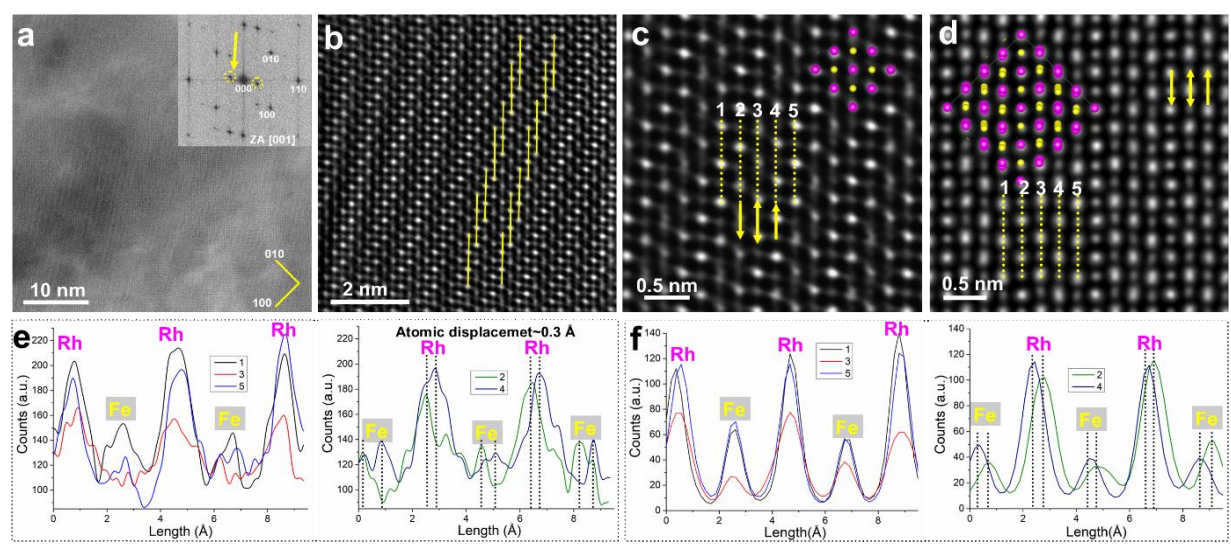

**Figure 3. a-c** HAADF-STEM images of FeRh alloy along [001] at different scales. Inset in **a** is an FFT pattern of **a**. **d** Simulated HAADF-STEM image using the proposed supercell. **e** is the line profiles for lines 1-5 in **c**. **f** is the line profiles for lines 1-5 in **d**. Arrows in **c** and **d** show direction of atomic displacement. The onset of the axis in **e** and **f** are where the line numbers are written.

To study the atom displacements from different viewing angles, HAADF-STEM studies were carried out on FeRh grains with [-111] zone axis orientation. Figure 4a and the corresponding FFT pattern (inset) confirms that the modulation is also present in this direction. Figure 4b shows a high-resolution HAADF-STEM image of the same region. Similar to what was observed in Figure 3a, the modulation seems to be caused by smearing of some atomic columns (yellow small arrows Figure 4b). The inset in Figure 4b, which is the line profile along dashed-yellow arrow, also reveals that the intensities of some atomic columns are lower than others. This can be attributed to atomic displacement in (110) plane along [1-10]. In order to see if the displacement matches the observations in [-111] as it was previously suggested, a simulated HAADF-STEM image of the proposed structure model along [-111] was prepared. However, a perfect match could not be found as it seemed that the displacement of atoms in (110) plane along [1-10] is not enough to reproduce the observed nanostructure along [-111]. Hence, the previously prepared supercell was taken as basis and a new supercell was prepared in which the atoms are not only displaced in (110) plane along [1-10], but also in (100) plane along [-100]. The schematics of this new supercell in [001] and [-111] zone axes are shown in Figure S10 and S11, respectively. The related cif (CIF FILE 1) file is attached to this manuscript as Supplementary Material 2. The HAADF-STEM image simulation and the simulated diffraction pattern of newly prepared supercell along [-111] are shown in Figure 4c and Figure S12, respectively. From these results, it can be seen that the HAADF-STEM observation and the simulated images match closely. Such ordered displacement of atoms is expected to induce localized strain in the nanostructure. The strain can be visualized by applying the geometrical phase analysis (GPA) on HAADF-STEM images. The image that was used for GPA analysis is shown as Figure S13. The results of the GPA analysis that are presented in Figure 4e, clearly indicate that the strain accommodation ($\varepsilon_{xx}$ and $\varepsilon_{xx}$ particularly) in the nanostructure is well-aligned with the displacement modulation. Thus, it can be concluded that the observed modulation is caused by ordered atomic displacement in {110} and {100} planes and, as expected these modulations are correlated to strain accommodation in the structure. Such modulated structures are commonly observed in unary, binary, ternary alloys with BCC structures and is known to be a premartensite structure state as mentioned in the Introduction.

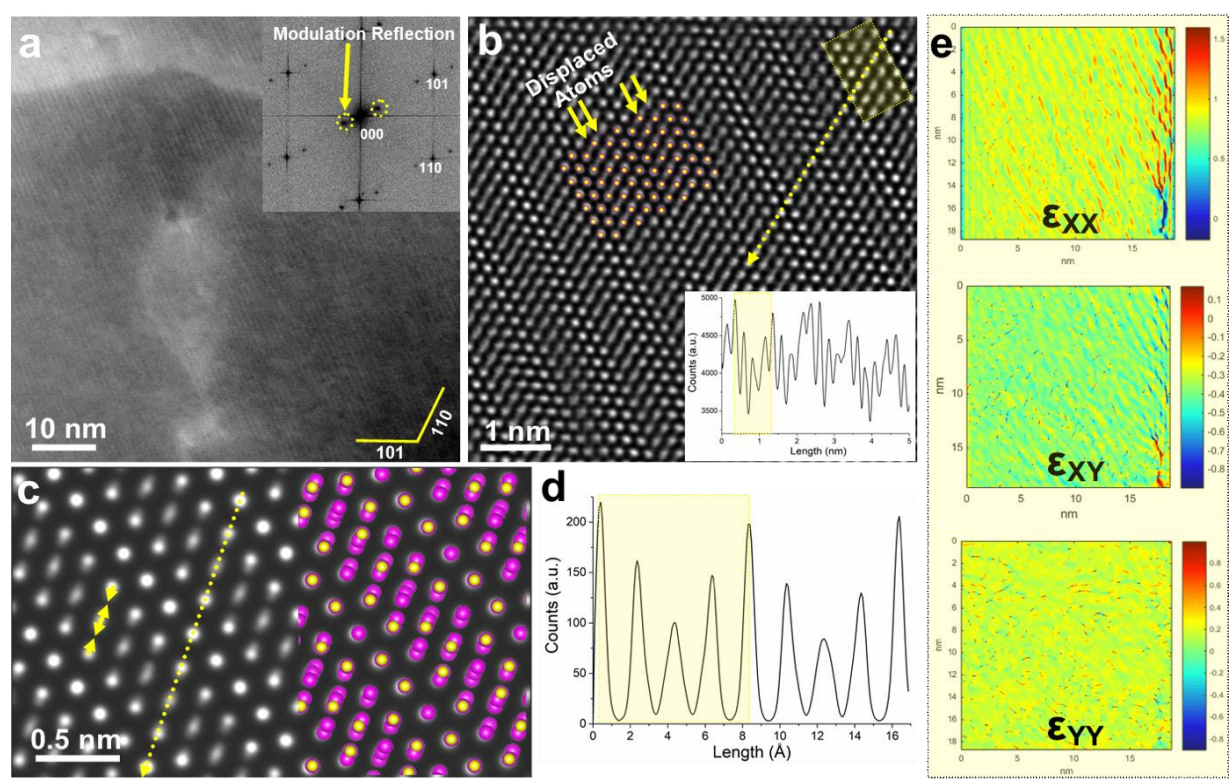

**Figure 4.** **a** and **b** HAADF-STEM images of FeRh alloy viewed along [-111] direction. Inset in **a** is an FFT pattern recorded from the same region. Inset in **b** is line profile along dashed arrow. **c** Simulated HAADF-STEM image

that partially overlapped with the proposed supercell (yellow Fe, purple Rh) and **d** is line profile along dashed arrow in **c**. Small arrows in **c** indicate atomic displacement direction. **e** Strain maps extracted from GPA analysis.

*(2) Quenching from 1150 °C into liquid nitrogen*

In order to investigate the effect of the thermal stress on the nanostructure of FeRh alloy and its magnetic properties, quenching in LN2 from 1150 °C was performed. Interestingly, as seen in Figure 5a, the magnetic properties of the alloy changed considerably after quenching in LN2, showing a large ferromagnetic background at low temperature. To correlate the ferromagnetic background with a possible phase transition, XRD and TEM examinations were carried out. The XRD pattern (Figure 5b) does not show formation of any new phase. In line with this, electron diffraction pattern of the alloy (Figure 5c) only shows presence of the B2-ordered structure. Yet, a closer look at the pattern reveals a new set of satellite reflections appearing at ~1/2{110} plane reflections, as shown by Figure 5d (See also Figure S14). From the DFT calculation, it was expected to observe extra reflections at such points, since very low frequency phonons were observed at points larger than $1/2TA2[\zeta\zeta 0]$. Figure 5e is an HRTEM image taken from the alloy (in [001] zone axis) which show nearly parallel domains in the nanostructure. Image filtering using fast Fourier transform (FFT) was carried out by selecting two of these 1/4{110} planes reflections (1 and 2 as indicated in Figure S14) which reveals that these domains are indeed areas with different displacement modulations. An overview TEM image of the domains in the same zone axis is given in Figure S15a, indicating that two 1/4(110) planes reflections (1 and 2 in Figure S14) are perpendicular to two 1/4(1-10) planes reflections (3 and 4 in Figure S14). The lamella was also examined in [0-11] zone axis and domain-like nanostructure was also seen as shown in Figure S15b. It is worth mentioning that such domain-like nanostructure was not reported for FeRh alloys previously.

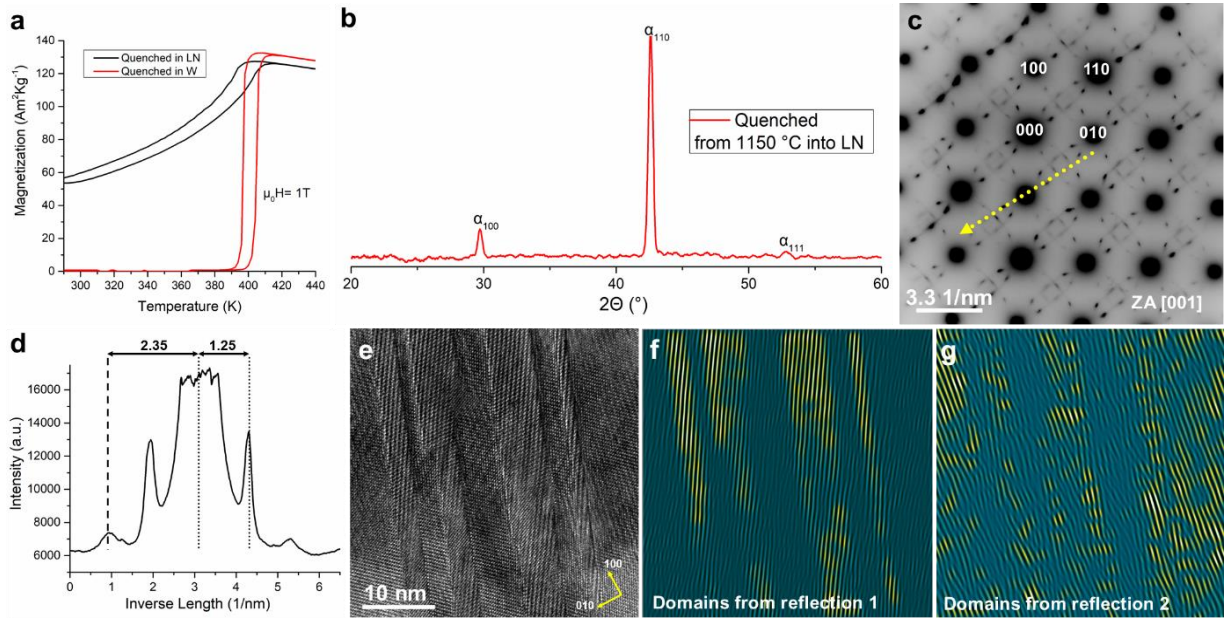

**Figure 5.** **a** Magnetization *vs.* temperature curve for FeRh alloy quenched from 1150 °C into water (red) and LN2 (black). **b** and **c** XRD and electron diffraction patern ([001]) of the alloy quenched in LN2. **d** Line profile recorded along dashed arrow in **c**. **e** HRTEM image of the alloy quenched in LN2. **f** and **g** Filtered FFT images obtained using satellite reflections (1 and 2) in Figure S14.

Regarding the 1/2{110} planes reflections, HAADF-STEM imaging along [001] zone axis was performed to reveal the cause. Figures 6a and b show HAADF-STEM image and corresponding FFT pattern, respectively. The later shows the presence of 1/2{110} planes reflections, while in the former the modulated structure is not easily visible. Similar to Figure 2, line profile analysis was done as shown in Figures 6c and d for line 1-4 in Figure 6a. Profiles of lines 1 and 3 are perfectly matching, while the same is true for line profiles of lines 2 and 4. The similarity of line profiles in 1 and 3, and 2 and 4 therefore leads to appearance of 1/2{110} planes reflections. The lower count intensity observed in the line profiles is considered to be related to atomic displacement, similar to that presented in Figures 3 and 4. The data show that by quenching sample in LN2, the length of modulation in real space decreased from four *d*-spacing of (110) planes (8.48 Å) to two (4.24 Å). However, it should be mentioned that these modulations are found to be incommensurate, so the modulation length change slightly at different places in the nanostructure. This is the reason for the 1/2{110} planes reflections in Figure 5c appearing streaked. Additionally, quenching in LN2 from 1150 °C resulted in more thermal stress leading to formation of a denser network of slightly different domains (areas in which atoms are displaced slightly in different orientation). Overall, it can be inferred that quenching FeRh alloy from 1150 °C into LN2 causes appearance of premartensite structure, in which portions of atoms are slightly displaced.

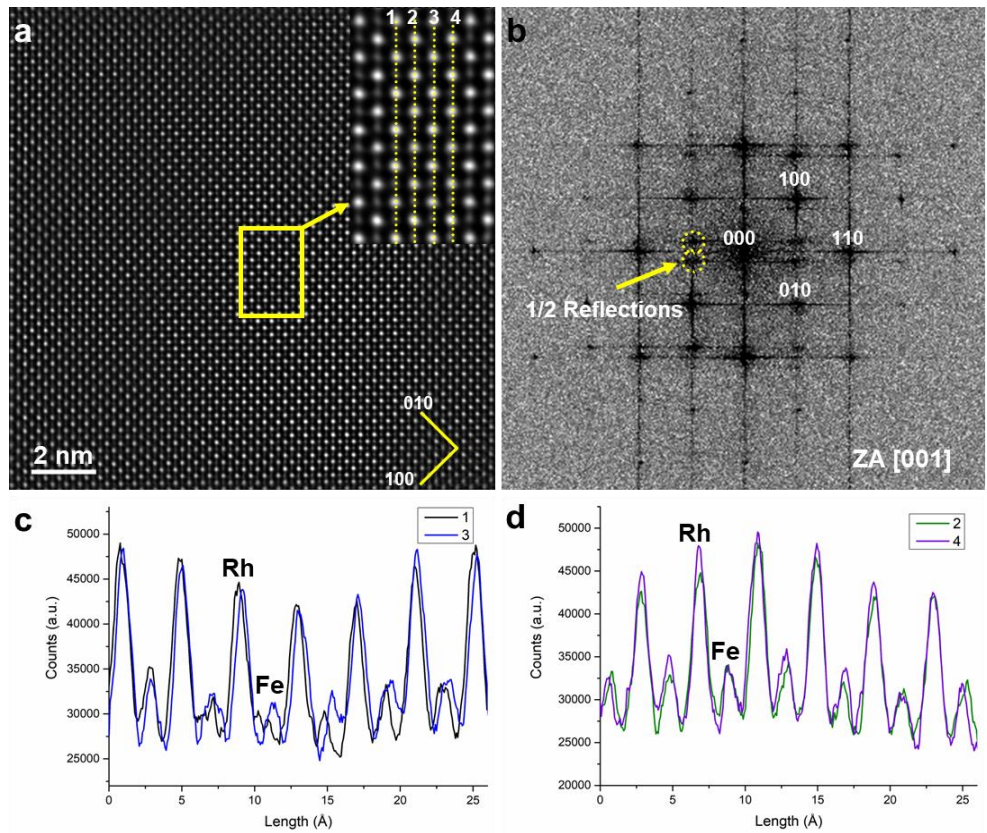

**Figure 6.** **a** HAADF-STEM image of areas with 1/2{110} planes reflections, which can be visibly seen in **b** corresponding FFT image. Inset in **a** is zoomed-in view of yellow rectangle. Line profiles (1-4 in **a**) are given in **c** and **d**.

*(3) Quenching from 1250 °C into liquid nitrogen*

With the hindsight of the effect of quenching in LN2 from 1150 °C on the alloy, an additional quenching experiment in LN2 was done, but from 1250 °C (at this temperature the alloy has still B2 structure [49]). Figure 7a shows the XRD measurements of the quenched alloy (black curve). Although the major phase is still B2, additional low intensity but broad peaks are observed at angles ~41°, 45° and 48°. According to previous reports on mechanical deformation of FeRh alloys [12,30,50,51], these peaks belong to the tetragonal L10 (a= 3.33 Å and c= 2.88 Å) and γ (a= 3.73 Å) phases, respectively. However, a considerable broadening is observed in the peaks, so a distribution of *d*-spacing is expected. In order to compare the effect of thermal stress with that of mechanical stress on FeRh system, a piece from the same batch of alloy was ~30 % compressed (stress-strain curve in Figure S16) and the obtained XRD pattern is shown in Figure 7a (red curve). Interestingly, mechanical stressing also results in formation of the same phases. To clarify the nanostructure of the sample quenched in LN2, a lamella was prepared using FIB. While preparing the lamella, a two-phase microstructure was observed when the thickness of lamella reached less than one micrometre (Figure 7b). When lamella was examined in bright-field TEM mode, a two-phase nanostructure was confirmed (Figure 7c). The diffraction pattern of the dark phase in Figure 7c was obtained (Figure 7c) and it was realized that although it seems to be γ, but in fact it is a L10 structure (tetragonal with *c/a* ratio of 1.4) that is modulated (see Figure S17) and has short range order (SRO). It was observed that the 100 and 001 reflections are split as seen more clearly in the low-magnification HAADF-STEM image and corresponding FFT pattern in Figures S18a and b. The high-magnification HAADF-STEM image and corresponding line profile are provided as Figures 7e and f. From these data, *c* and *a* are ~3.84 Å and ~2.78 Å, respectively. It is seen that the order is broken at some boundaries that are basically anti-phase boundaries. In fact, the anti-phase boundaries are the cause of the modulated structure in this phase. It was also observed that there is another L10 phase with lower *c/a* (~1.15), the DP of which is shown as Figure S19. The splitting of 100 and 001 reflections also occurs in this phase and the nanostructure is modulated too. It should be pointed out that these two different L10 structures are found to be in the same area with very coherent interfaces. Furthermore, other similar tetragonal structures (less abundant) with slightly different *c/a* were also observed in the present research that describes the broadness of the XRD peaks. However, since the most abundant tetragonal structures had *c/a* ratios of ~1.15 and 1.4, the discussion is continued only on these two phases. For simplification, now the L10 structures are called L10l and L10s, l and s standing for longer and shorter *c*, respectively. With the current data, now it is known that the previously-thought disordered γ phase is in fact L10l structure with SRO. The same is also true for L10s and therefore the SRO caused the disappearance of 100 and 001 peaks in XRD pattern of the current samples or those in the previous researches [29,51]. Overall, from these observations, it is concluded that imposing stress on FeRh B2 structure causes its transformation to L10 tetragonal structures with various *c/a* ratio.

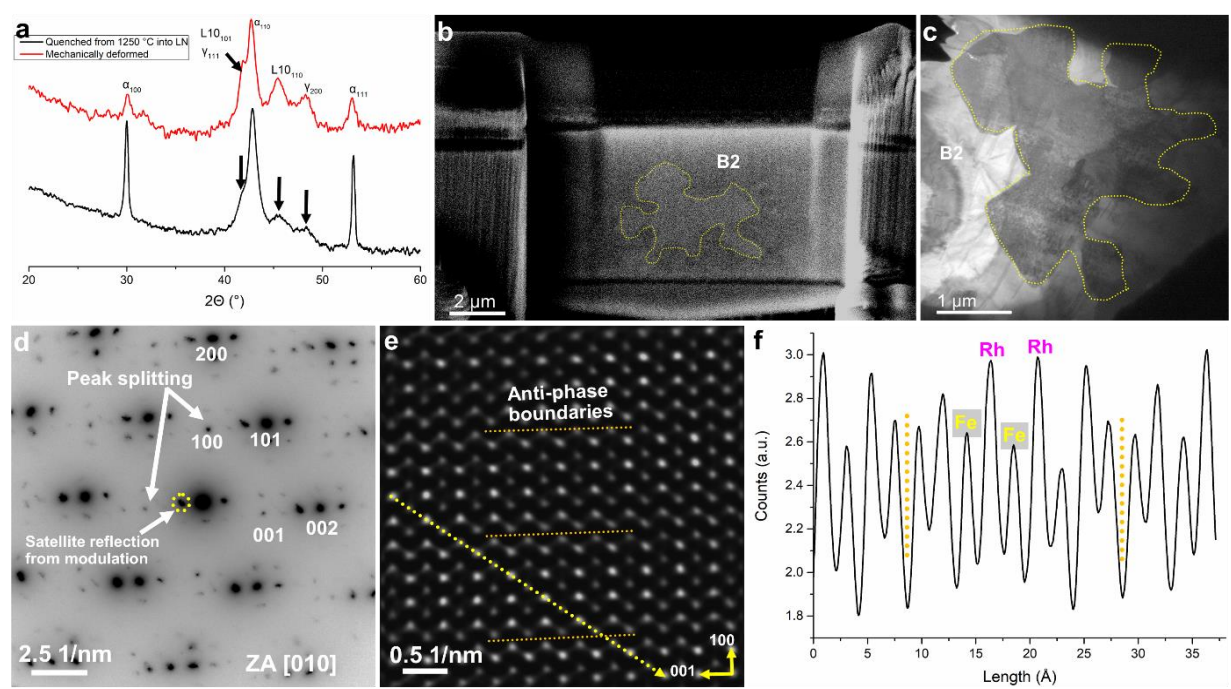

**Figure 7. a** XRD measurement of FeRh alloy quenched in LN2 from 1250 °C (black) and alloy mechanically deformed (red). **b** Secondary electron image of FIB lamella. **c** Bright-field TEM image taken from alloy quenched from 1250 °C into LN2. **d** Electron diffraction pattern and **e** HAADF-STEM image of L10l phase viewed along the [010] zone axis. **f** line profile taken along the dashed arrow in **e**.

In order to understand the mechanism of B2 transformation to these tetragonal structures an electron diffraction pattern was acquired from the interface of L10l and B2 near [001] zone axis of the latter as shown in Figure S20. As seen, the (-110) plane of B2 is parallel to (101) plane of L10, while (-101) and (002) form at positions of ~1/2{110} plane reflections, indicating that the transformation is displacive. Based on this orientation relationship, and according to the displacement mode that was observed in Figures 3 and 4, an atomic shuffling model that describes the transformation of B2 to L10 structures is proposed. Figure 8a shows the B2 structure along [001], while Figures 7b and c present the same structure while the atoms are shuffled according to our model to obtain L10l and L10s, respectively. Such atomic shuffling turns the B2 to L10l ([010]) and L10s ([010]). The developed model requires shuffling of atoms along one of the {110} planes in opposite or same directions and with different displacement length. The size and direction of arrow determine the length and direction of atomic shuffling. Resultant simulated diffraction patterns of B2, L10l, and L10s are shown as Figures 8d, e, and f, respectively. Additionally, superimposition of calculated diffraction pattern of B2 and L10l are shown as Figure S21. When this is compared with experimental diffraction pattern in Figure S20 a close match can be observed, indicating that our model describes the transformation very well.

The structures obtained in Figure 8e and f, although being very close to L10 structures, but have to be further adjusted in order to exactly resemble the L10 structures. The $c$ and $a$ lattice constants in Figure 5a are ~3.80 Å and ~2.36 Å, respectively. While the out of plane spacing (axis $b$) is 2.98 Å, same as B2 structure. Since $c$ is very close to what was experimentally observed (3.84 Å) in L10l, therefore no significant adjustment is required. However, in order to adjust $a$, the lattice should expand for ~16.5 % along [120] in reference to B2 structure as shown in Figure S22. Also, for the *b-axis* to shrink from 2.98 Å to 2.75 Å, there needs to be ~3.6 % shrinkage. This requires atoms to move along the out of plane direction ([001] in B2 structure). These movements are collective, meaning that all atoms are involved. For L10s (Figure 8f), the $a$ and $c$ are 3.39 Å and 2.65 Å, respectively. Again, $c$ is very close to what was observed in experiment (3.33 Å), so no significant adjustment is required. However, in order to get $a$ value to 2.88 Å, there needs to be 0.08 % expansion along [430]. For $b$ in L10s, same adjustment should occur as described for L10l.

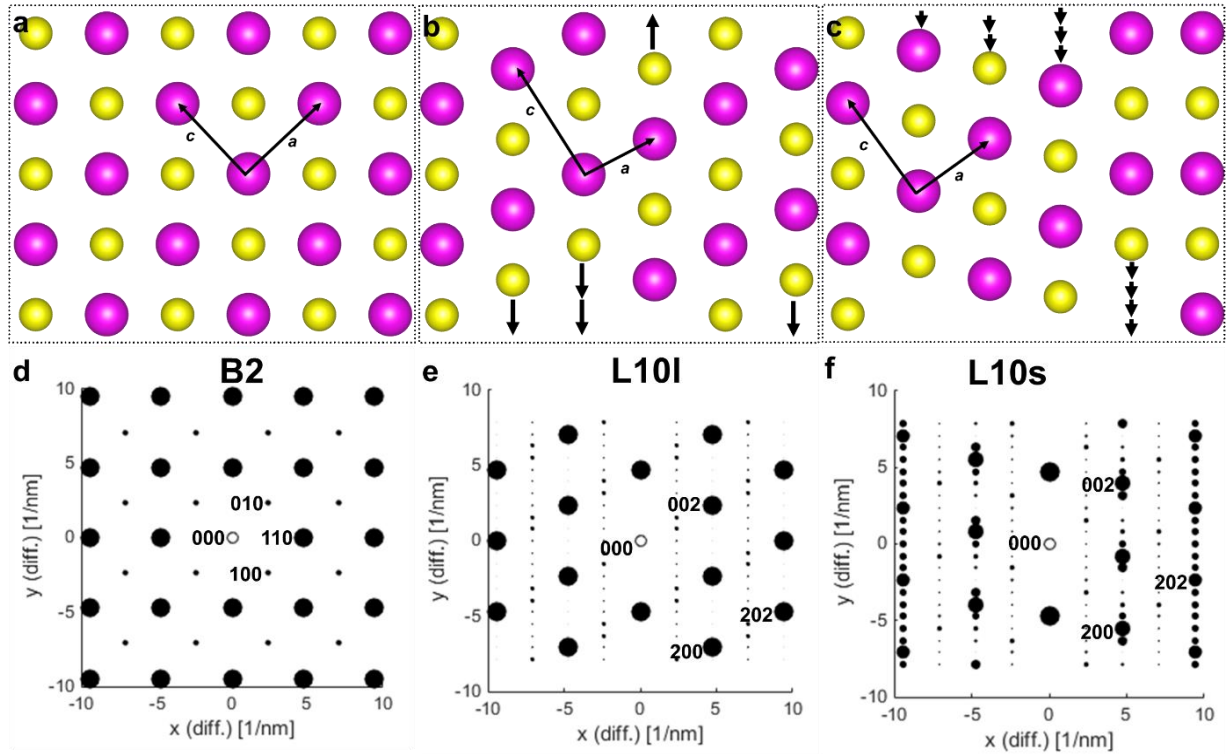

**Figure 8. a**, **b**, and **c** Atomic models for B2, L10l, and L10s structures, respectively. Size of arrows in **b** and **c** show value and direction of atomic displacement. Larger arrow length value is 1.06 Å, while that for shorter one is 0.53 Å. **d**, **e**, and **f** Calculated electron diffraction patterns of (B2), b (L10l), and c (L10s) models.

**Conclusion**

In the present study, a careful and comprehensive S/TEM study on thermally-stressed $Fe_{50}Rh_{50}$ alloy was done and following observations were made:

*(1)* For sample quenched in water from 1150 °C, it was observed that there is a systematic satellite reflection at ~1/4{110} and 1/4{100} planes reflections that are caused by slight displacement of Fe and Rh atoms in {110} and {100} planes along [1-10] and [-100], respectively. However, from phonon dispersion curve obtained by DFT satellite reflection at points larger than ~1/2{110} and 1/2{100} were expected.

*(2)* When sample was quenched from 1150 °C into LN2, 1/2{110} reflections appeared. With the assumption that the atomic structure of the alloy is frozen state of ~-167 °C, these reflections meet the expected phonon softening from obtained DFT calculation. It is believed that 1/2{110} and 1/4{110} are indications of the premartensitic structures. It should be also mentioned that quenching from 1150 °C into LN2 did not cause change in B2 structure, but considerably influenced the magnetic properties (broadening of AFM to FM transition).

*(3)* Quenching in LN2 from 1250 °C finally led to B2 phase transformation to two L10 structures with different *c/a* ratio (~1.15 and 1.4). The mechanically compressed alloy (~30 %) also showed formation of these two tetragonal phases, leading to conclusion that imposing external stress on FeRh alloy results in B2 transformation to tetragonal phases. TEM investigation of the L10 structures presented that both are short-range ordered and modulated. Modulations are caused by anti-phase boundaries, in which the order is broken. This explained the absence of 100 and 001 (ordering peaks) in the current and previous studies, leading researchers to conclude that γ disordered phase is the final martensitic product.

*(4)* From the S/TEM observations a model is proposed that describes the martensitic transformation of B2 to these tetragonal structures. Our model requires mainly shear mode of {110}<1-10>.

*(5)* According to the results of current study, it is concluded that B2 structure of FeRh in AFM states has intrinsic instability, same as other alloys with B2 structure. Upon exposure of external stress (thermal or mechanical) on FeRh, depending on the extent of the external field, it may transform to two tetragonal phases.

**Acknowledgment**

All authors acknowledge the financial support of German Science Foundation (DFG) in the framework of the Collaborative Research Centre Transregio 270 (CRC-TRR 270, project number 405553726, sub-projects Z01, Z02, A01, A10, B06). ZR and BG acknowledge support from the DFG through the award of the Leibniz Prize 2020. EA also acknowledge David Koch and Alex Aubert for his helpful discussion regarding sample preparation and XRD results.

**Declaration of Competing Interest**

The authors declare that they have no known competing financial interests of personal relationships that could have appeared to influence the work reported in this paper.

**Supplementary Materials 1:**

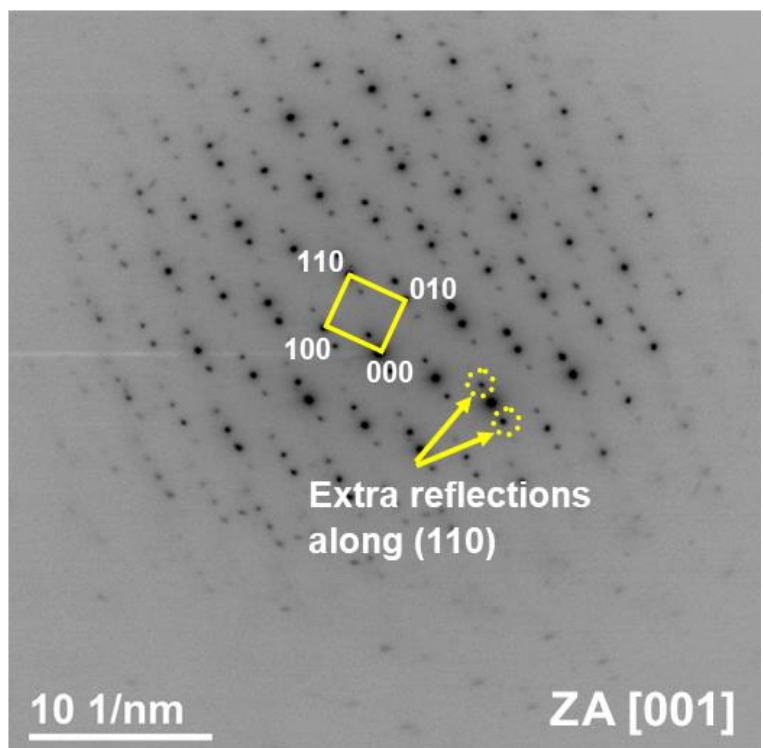

**Figure S1:** SAED pattern of FeRh viewed along [001] direction showing extra reflection along (110) plane.

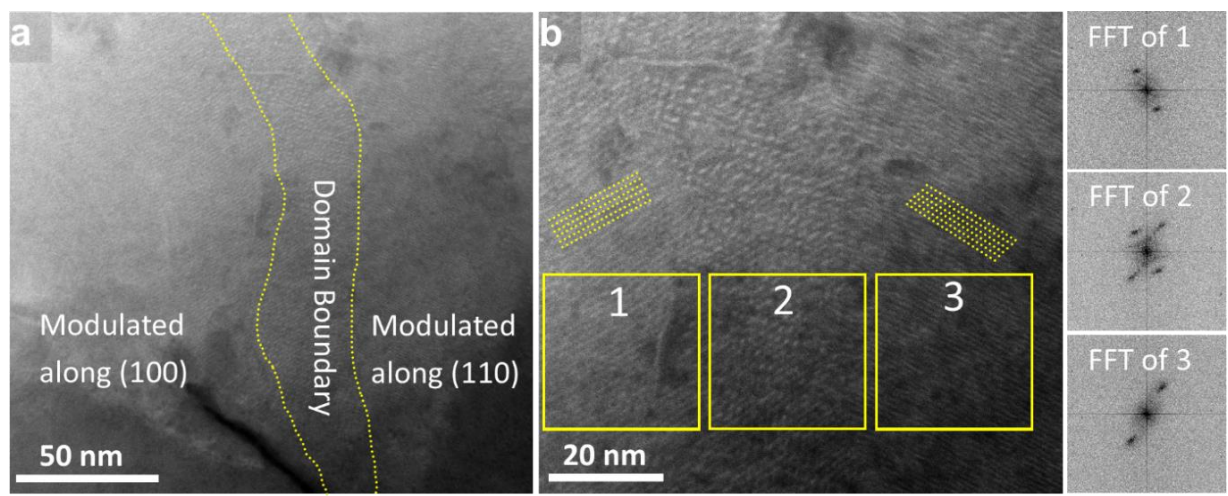

**Figure S2:** Domains with different modulations are shown in low-magnification HAADF-STEM images. Between domains there is domain boundary that has two modulations mixed, revealed by fast Fourier transform (FFT) images.

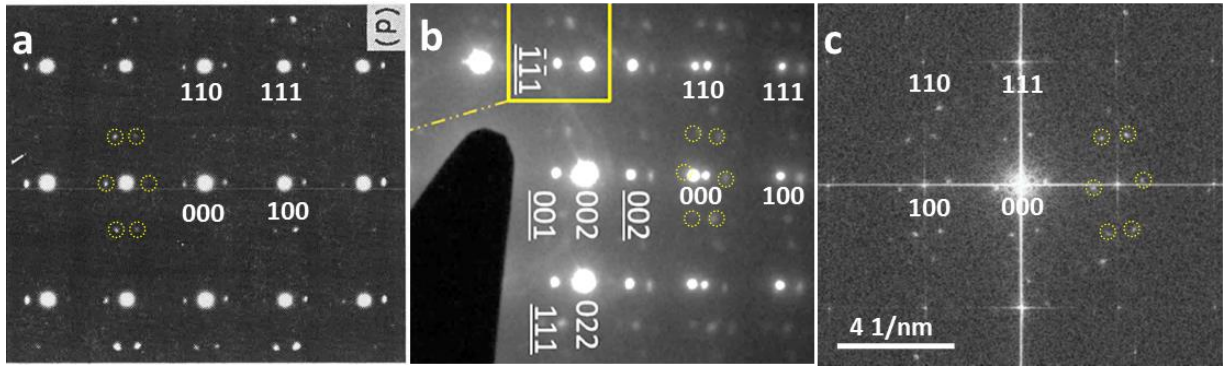

**Figure S3:** **a** Diffraction pattern (along [1-10]) taken from work done by Takahashi and co-workers [1] on cold rolled (30%) $Fe_{49.5}Rh_{50.5}$ alloy. **b** Diffraction patterns taken from work done by Castiella and co-workers [2] on thin film $Fe_{50}Rh_{50}$ (50 nm) grown on MgO single crystal annealed at 700 °C for 6 hours and **c** FFT pattern taken from our sample ($Fe_{50}Rh_{50}$) annealed for 72 hours at 1150 °C and quenched in water. The dashed-yellow pattern of circles in the diffraction patterns are the extra reflections.

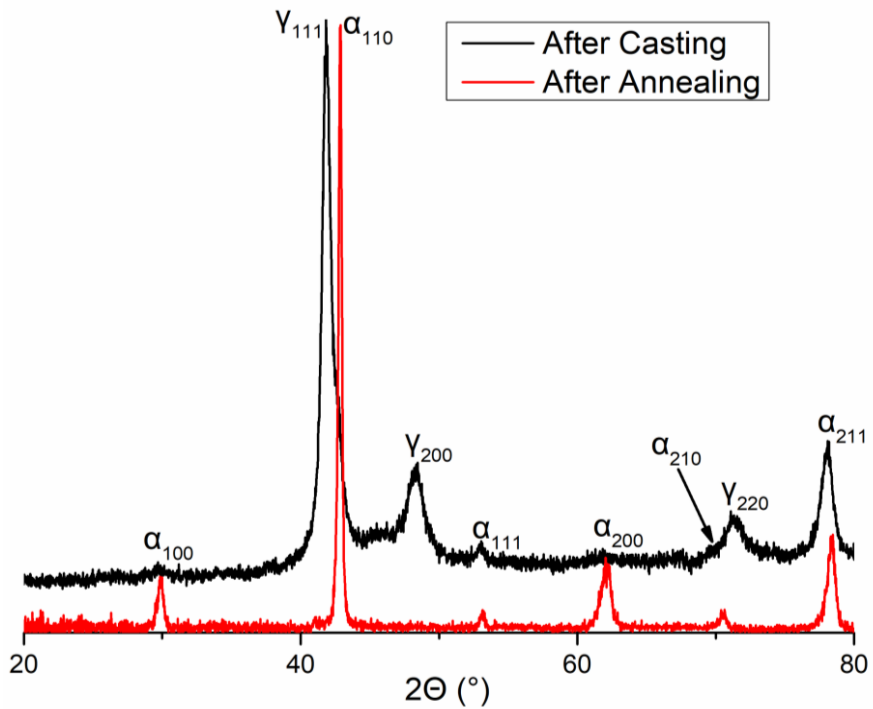

**Figure S4:** XRD measurements of $Fe_{50}Rh_{50}$ alloys: red is when sample is casted from the melt and black is for sample annealed at 1150 °C for 72 h and then quenched into water.

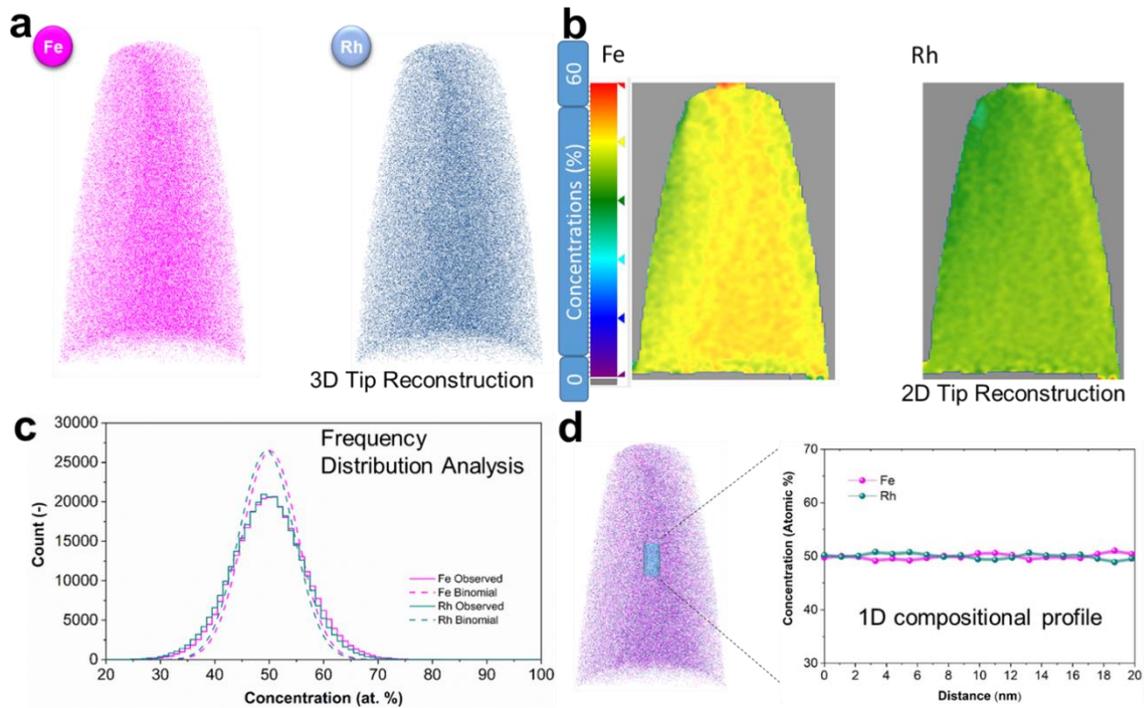

**Figure S5:** **a** APT reconstructions of Fe and Rh, **b** in-plane compositional analysis of Fe and Rh, **c** frequency distribution analysis obtained both from the observed experimental results and from the binomial simulation, **d** 1D compositional profile along the length direction of the cylinder shown in 3D APT tip.

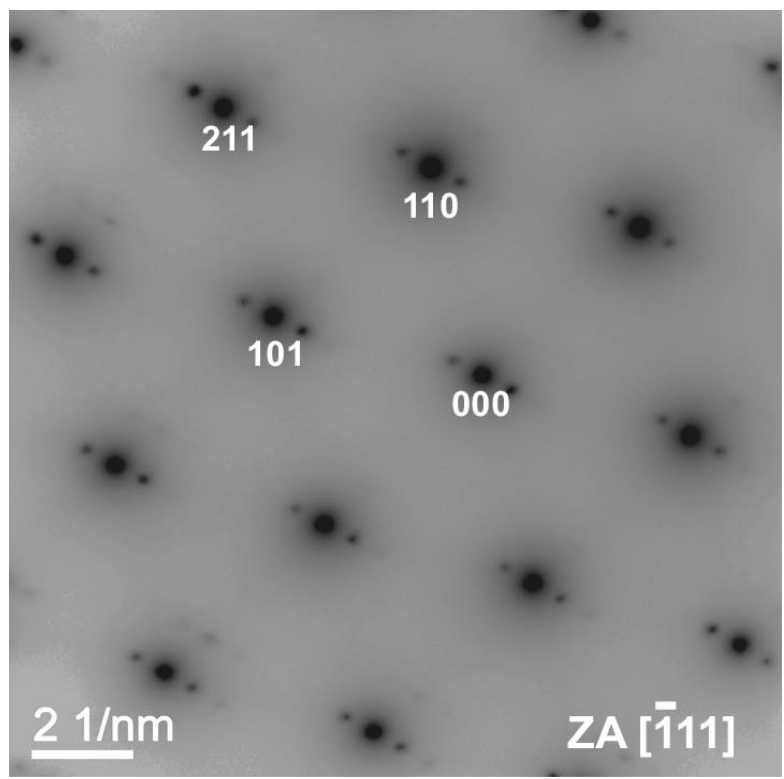

**Figure S6:** Diffraction pattern taken along [-111] direction from $Fe_{50}Rh_{50}$ alloy cooled with cooling rate of 0.1 degree/min.

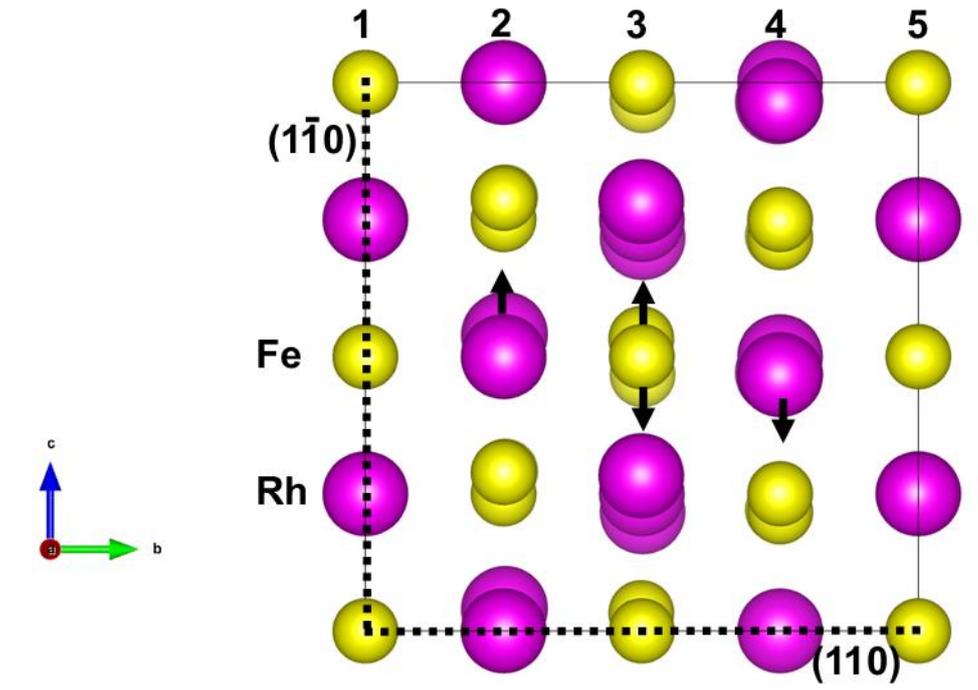

**Figure S7:** Prepared supercell of FeRh according to displacement observed in HAADF-STEM data. The arrows indicate direction of displacement. Atoms in lines 1 and 5 are not displaced. 50% of atoms in lines 2 and 4 are displaced. 50% of atoms in line 3 are also displaced. The displacement length is same, but the directions of displacement is different for various atoms.

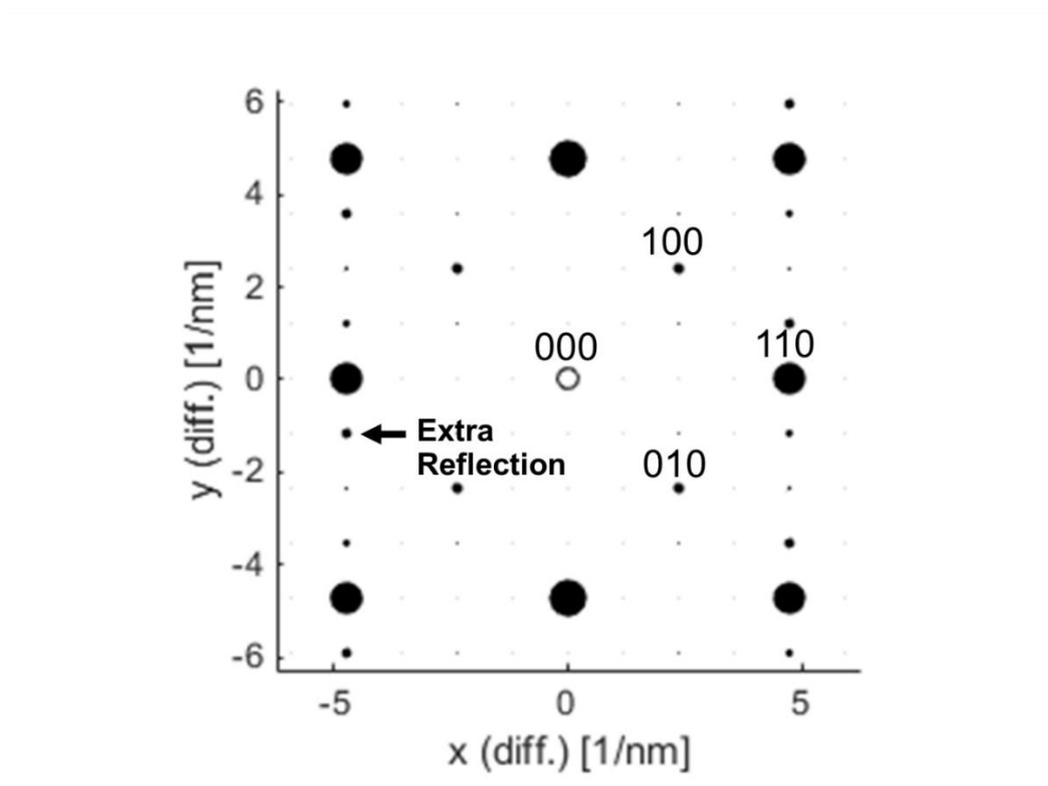

**Figure S8:** Simulated diffraction pattern from prepared supercell along [001].

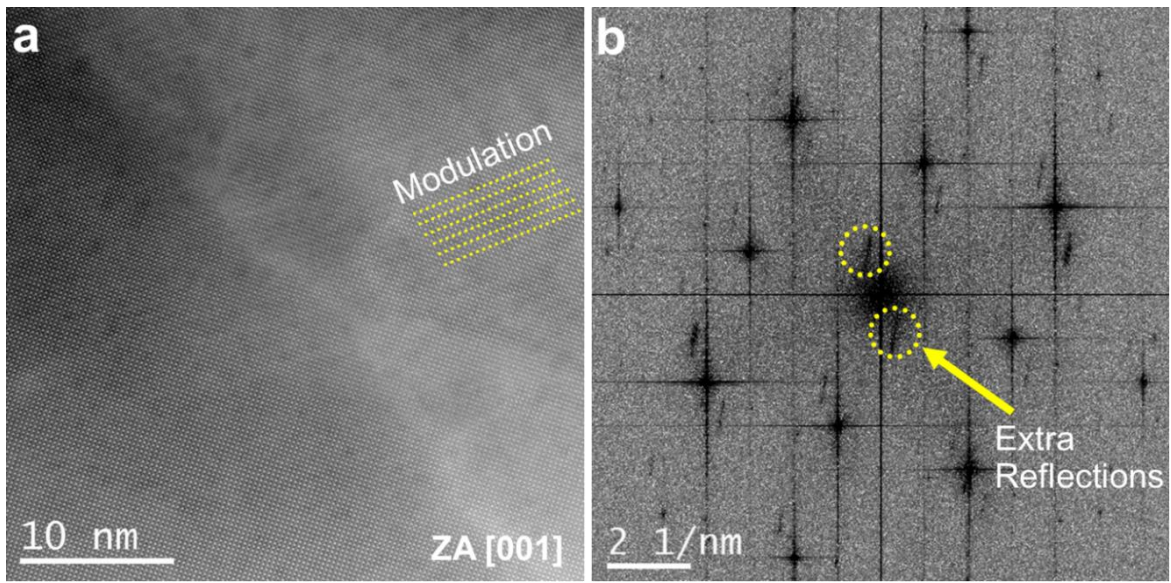

**Figure S9:** **a** HAADF-STEM image of FeRh viewed along [001] zone axis and **b** corresponding FFT pattern. The extra reflections are streaked, indicating that the modulation length is different at different locations even in a very small area. Along with change in the modulation length, the relative angle between the (110) reflection and the extra reflection is also changing.

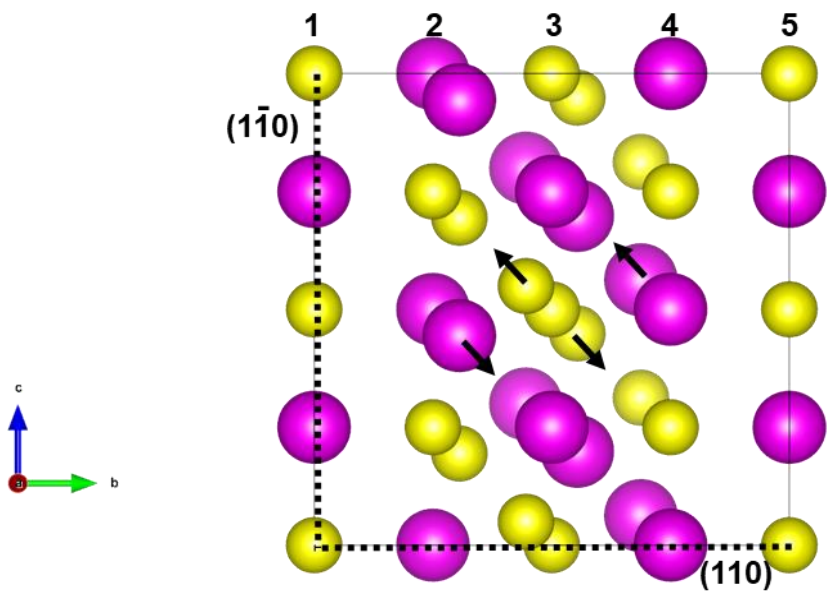

**Figure S10:** Prepared supercell according to displacement observed in HAADF-STEM data (from [001] and [-111] directions). The arrows indicate direction of displacement. Atoms in lines 1 and 5 are not displaced. 25% of atoms in lines 2 and 4 are displaced. 50% of atoms in line 3 are displaced. The displacement length is same, but the directions of displacement is different for various atoms.

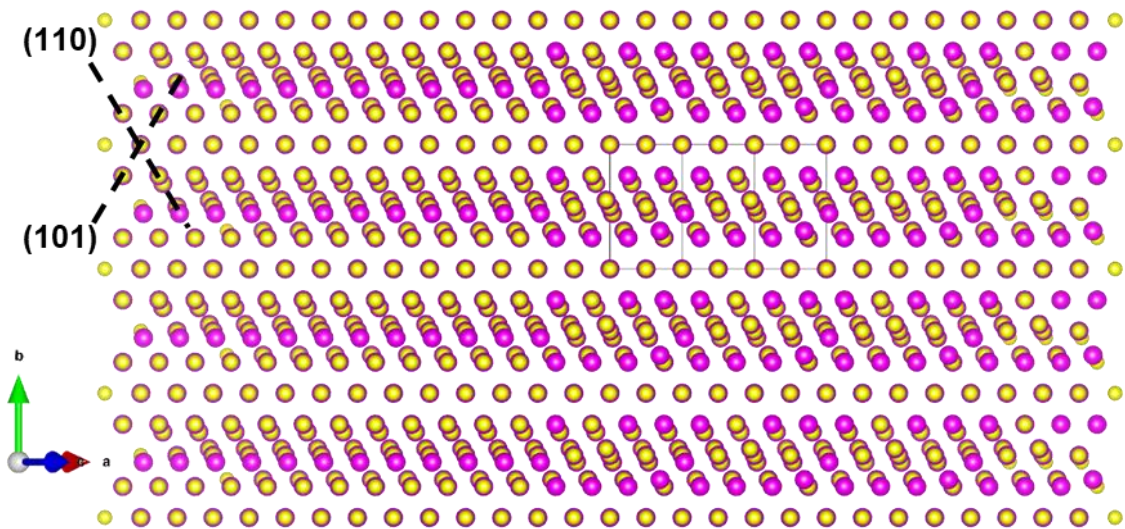

**Figure S11:** Prepared supercell of FeRh according to displacement observed in HAADF-STEM images recorded along [-111] direction. This figure is prepared by producing a 2×2×3 supercell of the supercell shown in Figure S10.

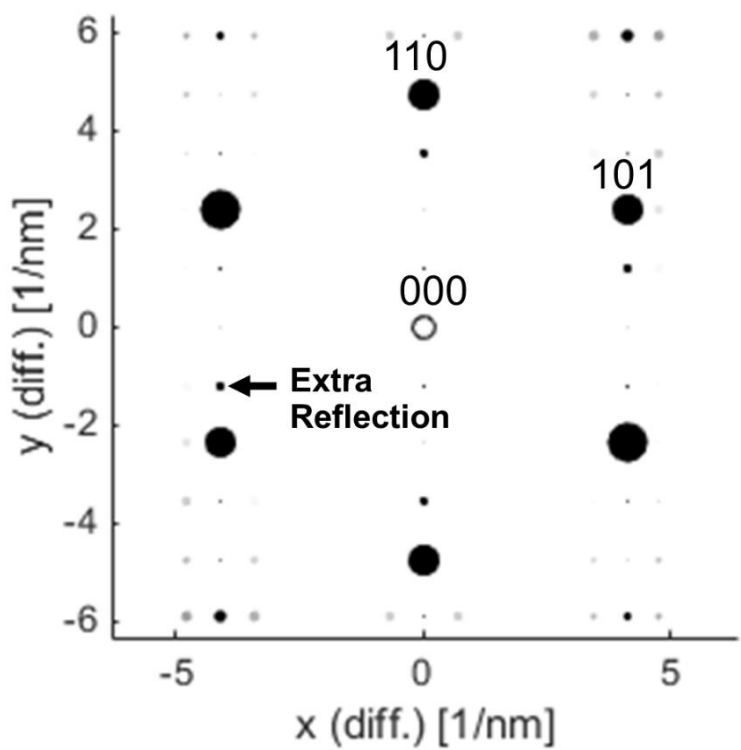

**Figure S12:** Simulated diffraction pattern along [-111] from prepared supercell according to displacement observed in HAADF-STEM data.

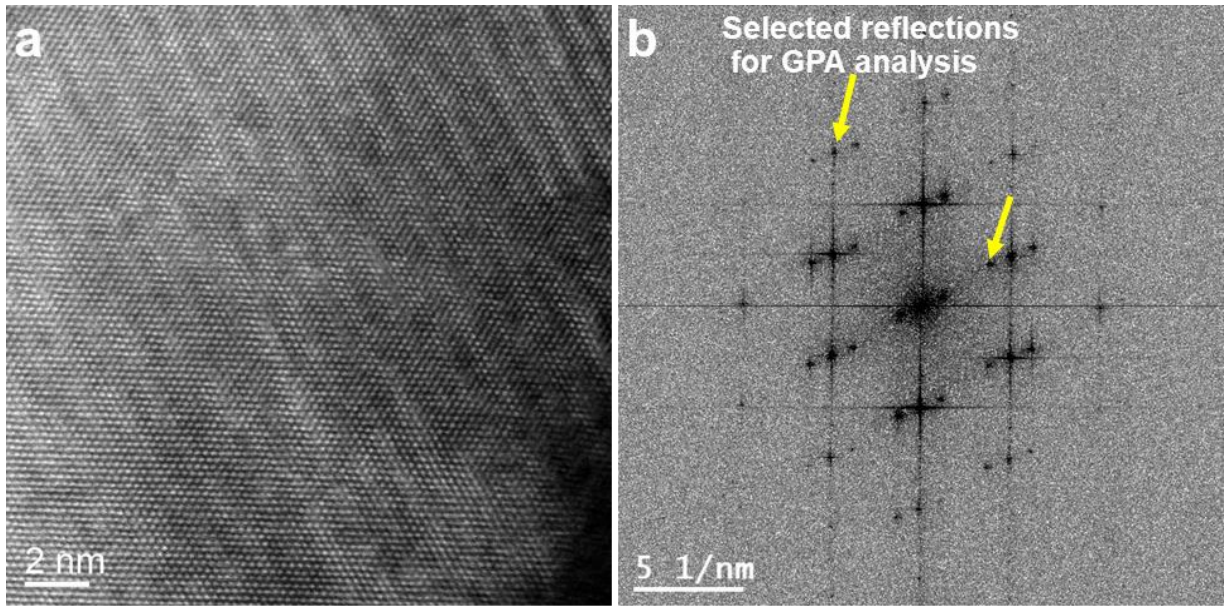

**Figure S13:** **a** HAADF-STEM image along [-111] and **b** corresponding FFT pattern used for GPA analysis. The reflections that were used for GPA analysis are indicated by yellow arrow in **b**.

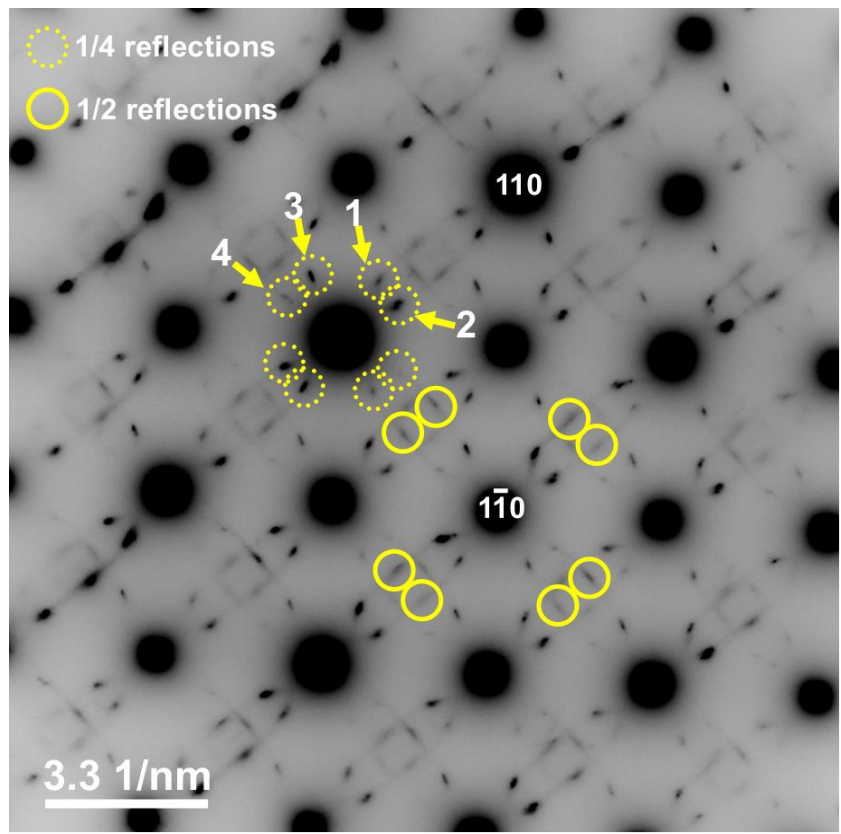

**Figure S14:** Electron diffraction pattern along [001] of FeRh alloy quenched in LN. In contrast to alloy quenched in water that showed only set of reflections at ~1/4{110} plane reflections, here we see another set of reflections at ~1/2{110} plane reflections. It should be noted that each of these extra reflections are coming from a domain. Appearance of two modulations at one domain was also observed occasionally.

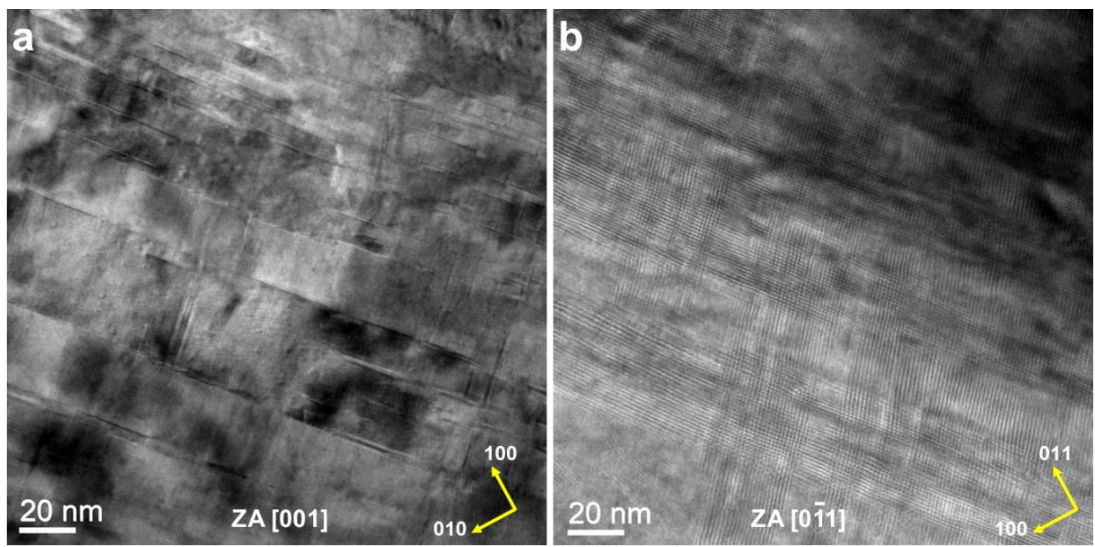

**Figure S15:** Low magnification TEM images of the FeRh alloy quenched in LN viewed along **a** [001] and **b** [0-11] crystallographic diractions.

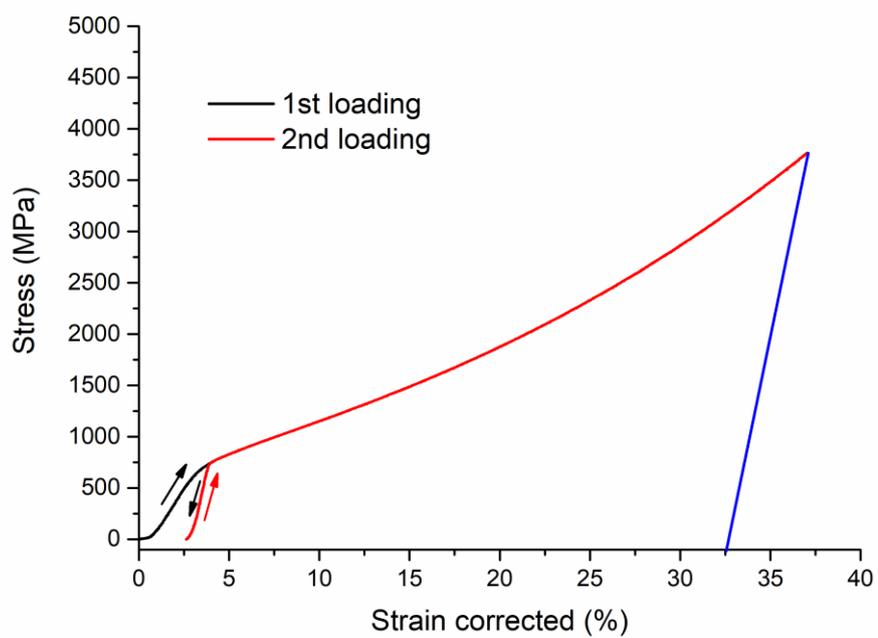

**Figure S16:** Stress-strain curve of FeRh alloy.

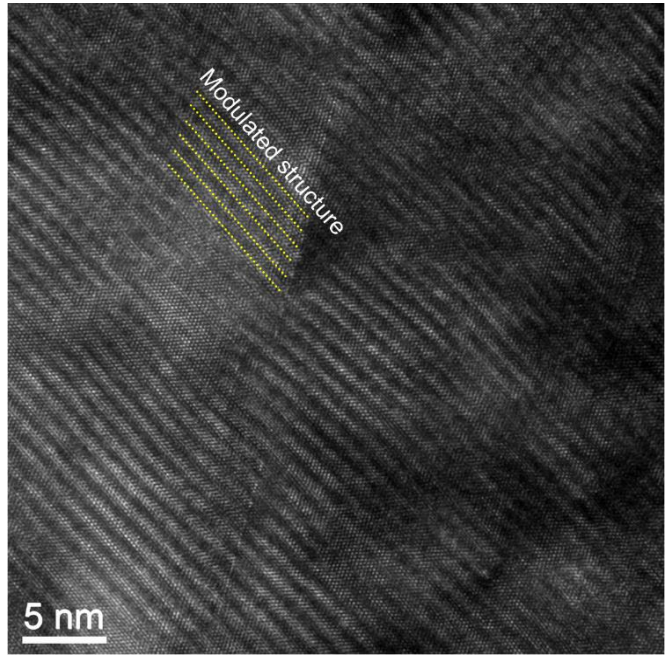

**Figure S17:** HRTEM image of L10 ($c/a$=1.4) phase in alloy quenched from 1250 °C into LN along [010].

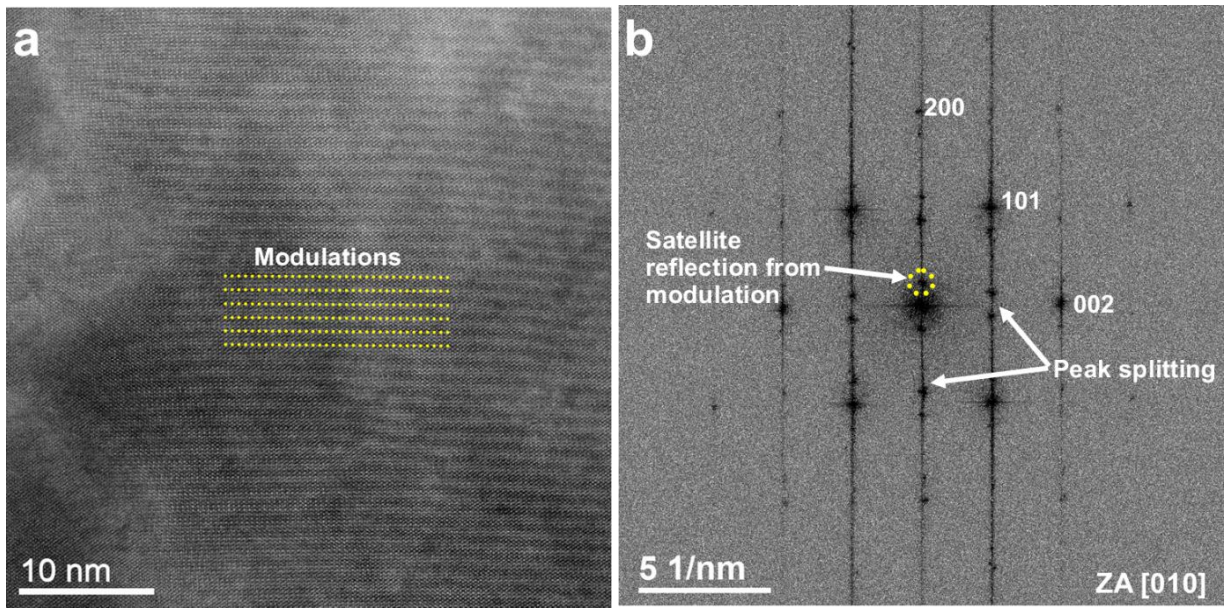

**Figure S18: a** HAADF-STEM image of L10 phase ($c/a$~1.4) in alloy quenched from 1250 °C into LN along [010] and **b** is corresponding FFT image.

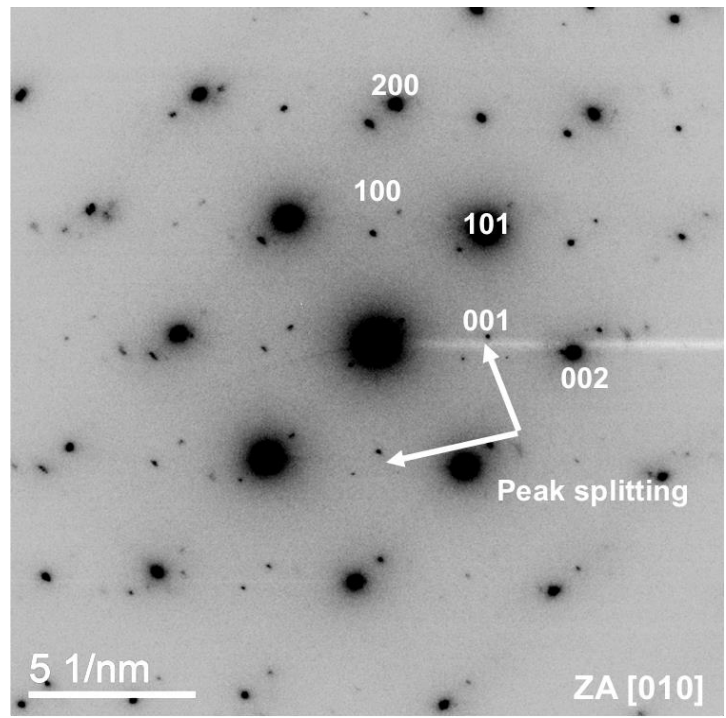

**Figure S19:** Electron diffraction pattern of L10 phase with a/c ratio of ~1.15.

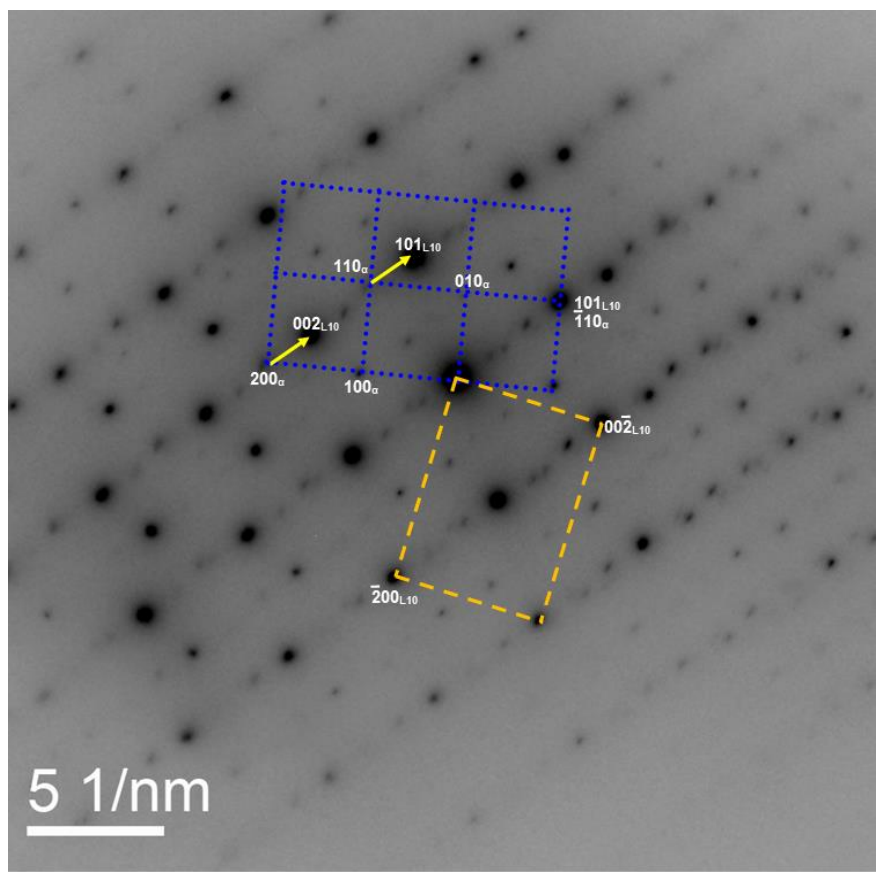

**Figure S20:** Electron diffraction pattern of L10l phase and B2 near [001] zone axis of the latter.

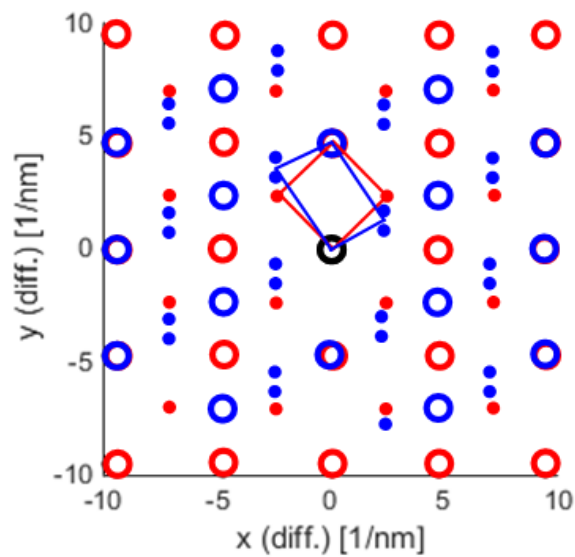

**Figure S21:** Superposition of calculated diffraction pattern of models of B2 (red) and L10l (blue).

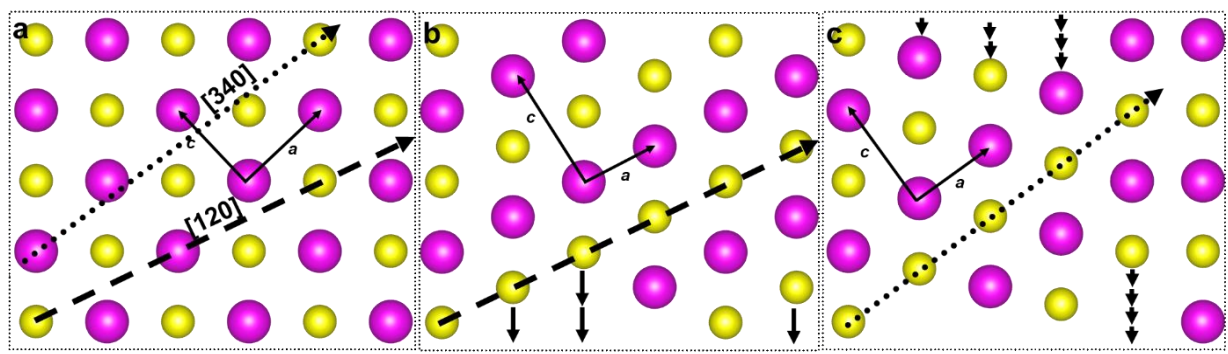

**Figure S22: a**, **b**, and **c** Atomic models for B2, L10l, and L10s, respectively. Further adjustments in *a* is required in order to obtain exact L10 structure. Here directions of atomic expansions (large dashed arrows) are shown for both L10l ([120]) and L10s ([340]) in regards to B2 reference state.

**Supplementary Materials 2:**

```
data_global
_cell_length_a 11.948
_cell_length_b 8.44851
_cell_length_c 8.44851
_cell_angle_alpha 90
_cell_angle_beta 90
_cell_angle_gamma 90
_symmetry_space_group_name_H-M 'P -1'
loop_
_symmetry_equiv_pos_as_xyz
 'x,y,z'
loop_
_atom_site_label
_atom_site_fract_x
_atom_site_fract_y
_atom_site_fract_z
Rh 0.125 6.57052e-17 0.25
Rh 0.125 6.57052e-17 0.75
Rh 0.117981 0.445614 0.299504
Rh 0.117981 0.445614 0.799504
Rh 0.375 6.57052e-17 0.25
Rh 0.375 6.57052e-17 0.75
Rh 0.375 0.5 0.25
Rh 0.375 0.5 0.75
Rh 0.625 6.57052e-17 0.25
Rh 0.625 6.57052e-17 0.75
Rh 0.624758 0.553809 0.198897
Rh 0.624758 0.553809 0.698897
Rh 0.875 6.57052e-17 0.25
Rh 0.875 6.57052e-17 0.75
Rh 0.875 0.5 0.25
```

Rh 0.875 0.5 0.75

Fe -2.6344e-17 0.25 0.25

Fe -2.6344e-17 0.25 0.75

Fe -2.6344e-17 0.75 0.25

Fe -2.6344e-17 0.75 0.75

Fe 0.25 0.25 0.25

Fe 0.25 0.25 0.75

Fe 0.245666 0.687571 0.312693

Fe 0.245666 0.687571 0.812693

Fe 0.5 0.25 0.25

Fe 0.5 0.25 0.75

Fe 0.5 0.75 0.25

Fe 0.5 0.75 0.75

Fe 0.75 0.305686 0.194314

Fe 0.75 0.305686 0.694314

Fe 0.75 0.75 0.25

Fe 0.75 0.75 0.75

Fe -1.55196e-17 1.05128e-16 0.5

Fe -1.55196e-17 1.05128e-16 1

Fe -0.00701856 0.445614 0.549504

Fe -0.00701856 0.445614 1.0495

Fe 0.25 1.05128e-16 0.5

Fe 0.25 1.05128e-16 1

Fe 0.25 0.5 0.5

Fe 0.25 0.5 1

Fe 0.5 1.05128e-16 0.5

Fe 0.5 1.05128e-16 1

Fe 0.499758 0.553809 0.448897

Fe 0.499758 0.553809 0.948897

Fe 0.75 1.05128e-16 0.5

Fe 0.75 1.05128e-16 1

Fe 0.75 0.5 0.5

Fe 0.75 0.5 1

Rh 0.125 0.25 0.5

Rh 0.125 0.25 1

Rh 0.125 0.75 0.5

Rh 0.125 0.75 1

Rh 0.375 0.25 0.5

Rh 0.375 0.25 1

Rh 0.370666 0.687571 0.562693

Rh 0.370666 0.687571 1.06269

Rh 0.625 0.25 0.5

Rh 0.625 0.25 1

Rh 0.625 0.75 0.5

Rh 0.625 0.75 1

Rh 0.875 0.305686 0.444314

Rh 0.875 0.305686 0.944314

Rh 0.875 0.75 0.5

Rh 0.875 0.75 1